\documentclass[11pt]{article}
\pdfoutput=1
\usepackage{array}
\usepackage{multirow}
\usepackage{amsmath}
\usepackage{amssymb}
\usepackage{mathtools}
\usepackage{pifont}
\usepackage{hhline}
\usepackage{graphicx}
\usepackage{cite}
\usepackage{xcolor}
\usepackage[colorlinks, linkcolor=red, anchorcolor=black,
citecolor=green]{hyperref}
\usepackage{longtable}
\usepackage{mathrsfs}
\usepackage{soul}

\sethlcolor{lightgray}

\newcolumntype{N}{@{}m{0pt}@{}}

\LTcapwidth=\textwidth

\begin{document}

\title{\hfill ~\\[-30mm] \hfill\mbox{\small USTC-ICTS-17-06}\\[10mm]
        \textbf{\Large Systematic Study of One-Loop Dirac Neutrino Masses and Viable Dark Matter Candidates}}

\date{}

\author{\\[1mm]Chang-Yuan Yao\footnote{Email: {\tt
phyman@mail.ustc.edu.cn}}~,~Gui-Jun Ding\footnote{Email: {\tt
dinggj@ustc.edu.cn}}\\ \\
\it{\small Interdisciplinary Center for Theoretical Study and  Department of
Modern Physics, }\\
\it{\small University of Science and
    Technology of China, Hefei, Anhui 230026, China}\\[4mm] }
\maketitle

\begin{abstract}

We study the generation of small neutrino masses from the dimension six effective operator at tree level and one-loop level, the neutrinos are assumed to be Dirac particles. We find out all possible tree and one-loop diagrams which lead to finite neutrino masses. The relevant mediator fields and the predictions for neutrino masses are presented.
The lower order contributions to neutrino masses can be forbidden
by introducing soft breaking abelian symmetry or finite non-abelian symmetry. The messengers inside the loop can be dark matter particles. The possible dark matter candidates and the corresponding constraints on the parameter $\alpha$ are analyzed for each model.

\end{abstract}
\thispagestyle{empty}
\vfill

\newpage
\setcounter{page}{1}
\section{Introduction}
The origin of the tiny but nonzero neutrino masses is an open question in
particle physics. A variety of mechanisms have been proposed to naturally
generate the neutrino masses. If neutrinos are Majorana particles, their
masses are uniquely described by the following effective operators in the
standard model (SM)~\cite{Babu:2001ex,Liao:2010ku},
\begin{equation}
\mathscr{L}_{5+2m}^M=-\frac{1}{2}\frac{g_{\alpha\beta}}{\Lambda}\left(\overline{\ell_{L\alpha}^C}\widetilde{H}^*\right)\left(\widetilde{H}^{\dagger}\ell_{L\beta}\right)\left(\frac{H^{\dagger}H}{\Lambda^2}\right)^m+\text{H.c.}\,,
\end{equation}
where $m$ is a generic non-negative integer, $\ell_L=(\nu_{L},l_{L})^T$ is
the left-handed lepton doublet, $H=(H^+,H^0)^T$ is the Higgs doublet with
$\widetilde{H}=i\sigma_2H^{*}$, $\Lambda$ denotes the new physics scale. The
famous Weinberg operator~\cite{Weinberg:1979sa} is exactly reproduced for
the case of $m=0$. The above neutrino mass operators can be obtained by
integrating out the new heavy fermions and scalars at tree level or through
radiative corrections at loop level. There are only three possible renormalizable ultraviolet completions of the Weinberg operators at tree level, and they are known as
type-I~\cite{Minkowski:1977sc,Yanagida:1979as,GellMann:1980vs,Mohapatra:1979ia},
type-II~\cite{Magg:1980ut,Schechter:1980gr,Wetterich:1981bx,Lazarides:1980nt,Mohapatra:1980yp,Cheng:1980qt},
and type-III~\cite{Foot:1988aq} seesaw mechanisms. In the scenario of radiatively generated neutrino masses, the effective
Weinberg operator can arise at one-loop~\cite{Zee:1980ai} or two-loop
level~\cite{Cheng:1980qt,Zee:1985id,Babu:1988ki}. A systematic analysis of
all possible one-loop realizations has been performed in
Ref.~\cite{Bonnet:2012kz}, and the possible dark matter particle as one of
the internal messengers is discussed in Ref.~\cite{Restrepo:2013aga}.
A similar analysis has been performed for the one-loop topologies
which lead to the dimension seven effective neutrino mass operator corresponding to $m=1$~\cite{Cepedello:2017eqf}. Systematic classifications of the two-loop radiative models are presented in Refs.~\cite{Sierra:2014rxa,Simoes:2017kqb}.
Radiative generation of neutrino mass is a very appealing idea, and there
are many such kind of models and relevant phenomenological studies in the
literature, see~\cite{Cai:2017jrq} for a recent review.

The signal of neutrinoless double beta decay has not been observed, the nature of neutrinos is still unclear, and we can not exclude the possibility that neutrinos are Dirac particles. In the context of standard model, the most general effective operators for the Dirac neutrino masses take the following form
\begin{equation}
\label{eq:lag_d4}\mathscr{L}_{4+2n}^D=-y_{\alpha\beta}\overline{\ell_{L\alpha}}\widetilde{H}\nu_{R\beta}\left(\frac{H^{\dagger}H}{\Lambda^2}\right)^n+\text{H.c.}\,,
\end{equation}
where $\nu_{R}$ are the right-handed neutrino fields. Since $\nu_{R}$ are
standard model singlets, the Majorana mass term
$(m_N/2)\overline{\nu_R^c}\nu_R$ is allowed such that the light neutrinos
would be Majorana particles. As a consequence, under the assumption that
neutrinos are Dirac particles, additional symmetry is generally needed to
forbid the right-handed neutrino Majorana mass term and it is usually taken
to be $U(1)_L$ lepton number. The lowest order dimensional four Dirac
neutrino mass operator for $n=0$ can be generated from the tree level
diagram in the so-called Dirac seesaw
mechanism~\cite{Roncadelli:1983ty,Roy:1983be,Gu:2006dc}. The Dirac neutrino masses can
also be realized through loop
corrections~\cite{Mohapatra:1987nx,Gu:2007ug,Kanemura:2011jj,Farzan:2012sa,Bonilla:2016diq,Ma:2017kgb}.
The generation of Dirac neutrino masses at tree level and one-loop level has
been studied in Refs.~\cite{Ma:2016mwh,Wang:2016lve}, and the possible topologies of the one-loop diagram are found.

In the present work, we shall focus on the dimension six operator of the
Dirac neutrino masses corresponding to $n=1$. We shall give a systematic
analysis of underlying ultraviolet completions of this operator at both
tree level and one-loop level. We shall identify all topologies and the
corresponding models in which the leading order contributions to the
neutrino masses arise from the dimension six operator while the dimension four one-loop contributions should be forbidden by the symmetry
of the models with a given field content. The resulting predictions of each viable model for the light neutrino mass matrix would be presented.

Another evidence for physics beyond the standard model is dark matter. The
dark matter plays an important role in cosmic structure formation and galaxy
formation and evolution and on explanations of the anisotropies observed in
the cosmic microwave background. The exact nature of dark matter is still
unknown. It is widely accepted that the dark matter particles must be
neutral and stable in order to be consistent with structure formation.
Various dark matter models have been put forward. There are many experiments that are searching for dark matter particles or are in various stages of planning and construction, but no dark matter
particle has been conclusively identified. Obviously both neutrino masses
and dark matter should be explained in the underlying model of new physics beyond the standard model. Analogous to the scotogenic model~\cite{Ma:2006km}, the particles mediating the one-loop diagram for neutrino masses could be dark matter candidates. In this work, we shall further study whether and under which conditions the above obtained dimension six Dirac neutrino mass models can account for dark matter.

This paper is organized as follows: In section~\ref{sec:diagrams} we
systematically study the tree level and one-loop realizations of the
dimension six Dirac neutrino mass operator, we identify all topologies and list the possible content of the mediators and their transformation properties under the SM gauge group for the viable models. In
section~\ref{sec:forbidtree}, we show how to forbid the lower order contributions to the neutrino masses. In section~\ref{sec:darkmatter}, we discuss the possible dark matter candidates
of the viable models. Finally, we summarize and present our conclusions in
section~\ref{sec:conclusion}. We show the one-loop Feynman diagrams, the possible assignments for the mediators, the expressions for the neutrino masses and the dark matter candidates in Appendix~\ref{sec:app_diagram}.
The explicit expressions of the loop integral involved in neutrino mass are collected in Appendix~\ref{sec:app_mass}.

\begin{table}[!htb]
\centering
\begin{small}
\begin{tabular}{|m{0.4cm}<{\centering}|m{1.4cm}<{\centering}|m{1.4cm}<{\centering}||m{0.4cm}<{\centering}|m{1.4cm}<{\centering}|m{1.4cm}<{\centering}||m{0.4cm}<{\centering}|m{1.4cm}<{\centering}|m{1.4cm}<{\centering}||m{0.4cm}<{\centering}|m{1.4cm}<{\centering}|m{1.4cm}<{\centering}|}\hline\hline
\multicolumn{3}{|m{4cm}<{\centering}||}{\multirow{-2}{*}{\includegraphics[width=\linewidth]{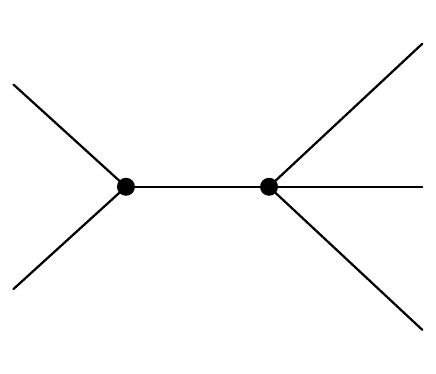}}} & \multicolumn{3}{m{4cm}<{\centering}||}{\multirow{-2}{*}{\includegraphics[width=\linewidth]{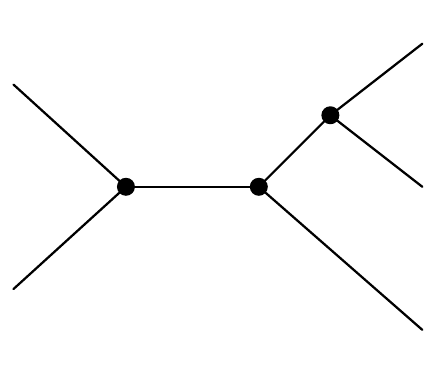}}} &  \multicolumn{3}{m{4cm}<{\centering}||}{\includegraphics[width=\linewidth]{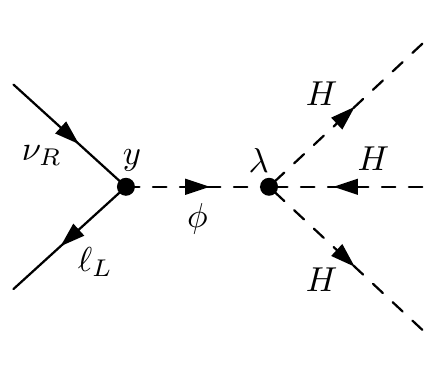}} &  \multicolumn{3}{m{4cm}<{\centering}|}{\includegraphics[width=\linewidth]{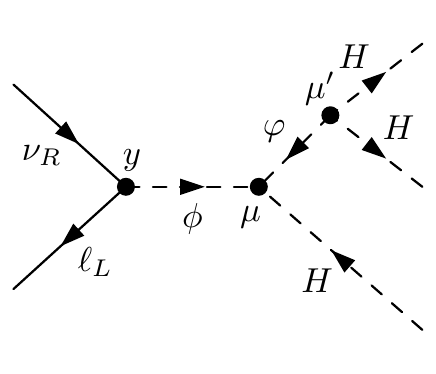}} \\[-0.5cm]
\multicolumn{3}{|c||}{} & \multicolumn{3}{c||}{} & \multicolumn{3}{c||}{F1-1-1} & \multicolumn{3}{c|}{F2-1-1} \\ \cline{7-12}
\multicolumn{3}{|l||}{} & \multicolumn{3}{l||}{} &  & \multicolumn{2}{c||}{$\phi$} &  & $\phi$ & $\varphi$   \\ \cline{7-12}
\multicolumn{3}{|c||}{F1} & \multicolumn{3}{c||}{F2} & A & \multicolumn{2}{c||}{$\mathbf{2}_{1}^S$} & A  & $\mathbf{2}_{1}^S$ & $\mathbf{3}_{-2}^S$  \\ \cline{7-12}
\multicolumn{3}{|l||}{} & \multicolumn{3}{l||}{} & \multicolumn{3}{c||}{$(m_{\nu})_{\alpha\beta}/\langle H\rangle^3=-\frac{\lambda y_{\alpha\beta}}{M_{\phi}^2}$} & \multicolumn{3}{c|}{$(m_{\nu})_{\alpha\beta}/\langle H\rangle^3=-\frac{\mu\mu' y_{\alpha\beta}}{M_{\phi}^2M_{\varphi}^2}$} \\ \hline\hline
\multicolumn{3}{|m{4cm}<{\centering}||}{\includegraphics[width=\linewidth]{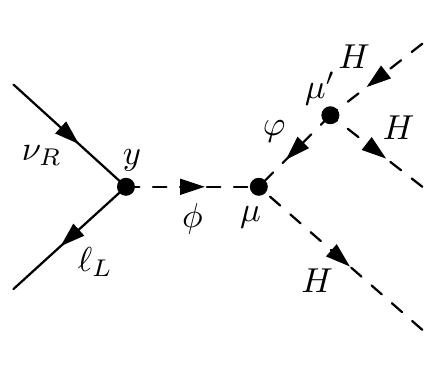}} &  \multicolumn{3}{m{4cm}<{\centering}||}{\includegraphics[width=\linewidth]{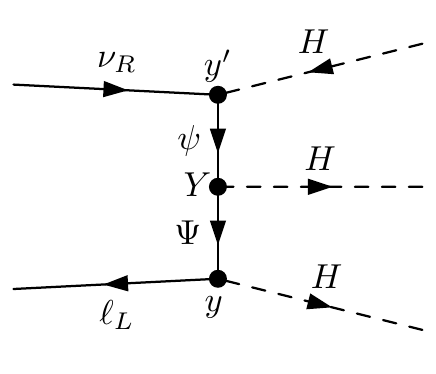}} &  \multicolumn{3}{m{4cm}<{\centering}||}{\includegraphics[width=\linewidth]{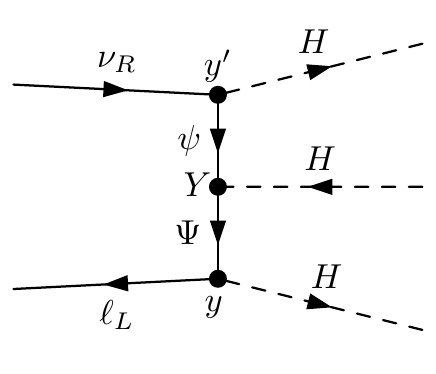}} &  \multicolumn{3}{m{4cm}<{\centering}|}{\includegraphics[width=\linewidth]{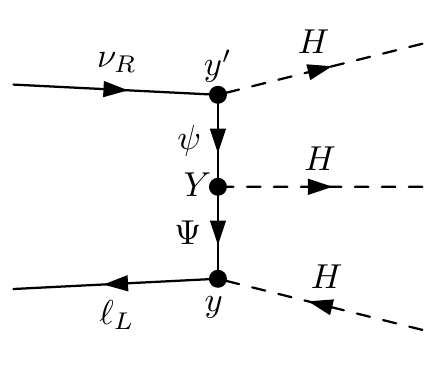}} \\ [-0.2cm]
\multicolumn{3}{|c||}{F2-1-2} & \multicolumn{3}{c||}{F2-2-1} & \multicolumn{3}{c||}{F2-2-2} & \multicolumn{3}{c|}{F2-2-3} \\ \hline
& $\phi$ & $\varphi$ & & $\psi$ & $\Psi$ & & $\psi$ & $\Psi$ & & $\psi$ & $\Psi$ \\ \hline
A & $\mathbf{2}_{1}^S$ & $\mathbf{1}_{0}^S$ & A & $\mathbf{2}_{1}^F$ & $\mathbf{1}_{0}^F$ & A & $\mathbf{2}_{-1}^F$ & $\mathbf{1}_{0}^F$ & \multirow{2}{*}{A} & \multirow{2}{*}{$\mathbf{2}_{-1}^F$} & \multirow{2}{*}{$\mathbf{3}_{-2}^F$}    \\ \cline{1-9}
B & $\mathbf{2}_{1}^S$ & $\mathbf{3}_{0}^S$ & B & $\mathbf{2}_{1}^F$ & $\mathbf{3}_{0}^F$ & B & $\mathbf{2}_{-1}^F$ & $\mathbf{3}_{0}^F$ &    &  &  \\ \hline
\multicolumn{3}{|c||}{$(m_{\nu})_{\alpha\beta}/\langle H\rangle^3=-\frac{\mu\mu' y_{\alpha\beta}}{M_{\phi}^2M_{\varphi}^2}$} & \multicolumn{3}{c||}{$(m_{\nu})_{\alpha\beta}/\langle H\rangle^3=-\frac{y_{\alpha i}Y_{ij}y'_{j\beta}}{M_{\Psi_i}M_{\psi_j}}$} & \multicolumn{3}{c||}{$(m_{\nu})_{\alpha\beta}/\langle H\rangle^3=-\frac{y_{\alpha i}Y_{ij}y'_{j\beta}}{M_{\Psi_i}M_{\psi_j}}$} & \multicolumn{3}{c|}{$(m_{\nu})_{\alpha\beta}/\langle H\rangle^3=-\frac{y_{\alpha i}Y_{ij}y'_{j\beta}}{M_{\Psi_i}M_{\psi_j}}$} \\ \hline\hline
\multicolumn{3}{|m{4cm}<{\centering}||}{\includegraphics[width=\linewidth]{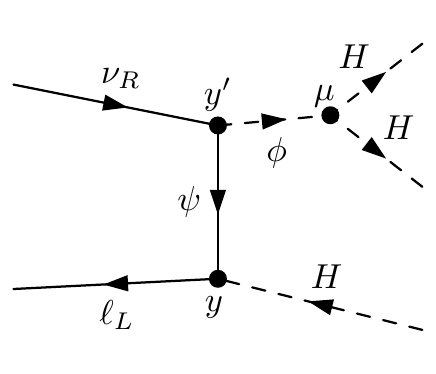}} &  \multicolumn{3}{m{4cm}<{\centering}||}{\includegraphics[width=\linewidth]{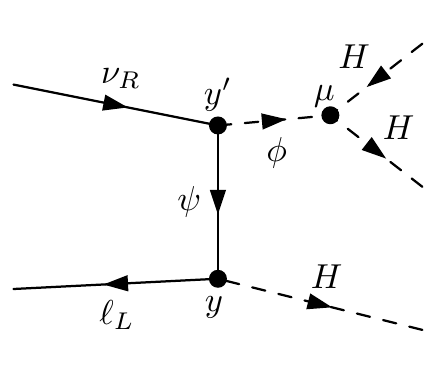}} &  \multicolumn{3}{m{4cm}<{\centering}||}{\includegraphics[width=\linewidth]{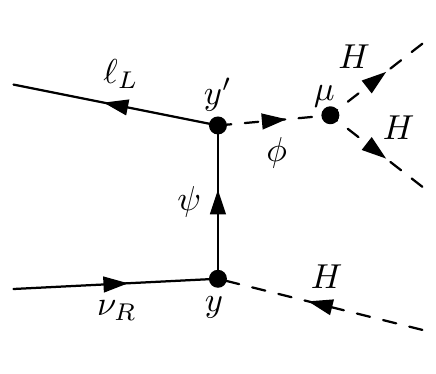}} &  \multicolumn{3}{m{4cm}<{\centering}|}{\includegraphics[width=\linewidth]{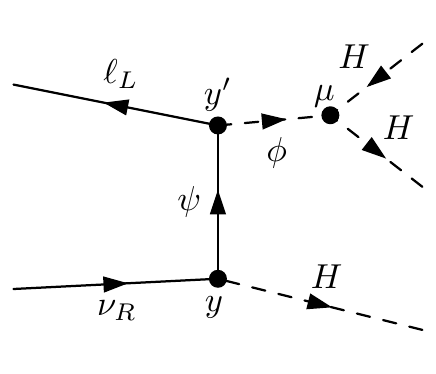}} \\ [-0.2cm]
\multicolumn{3}{|c||}{F2-3-1} & \multicolumn{3}{c||}{F2-3-2} & \multicolumn{3}{c||}{F2-3-3} & \multicolumn{3}{c|}{F2-3-4} \\ \hline
& $\psi$ & $\phi$ & & $\psi$ & $\phi$ & & $\psi$ & $\phi$ & & $\psi$ & $\phi$ \\ \hline
\multirow{2}{*}{A} & \multirow{2}{*}{$\mathbf{3}_{-2}^F$} & \multirow{2}{*}{$\mathbf{3}_{2}^S$} & A & $\mathbf{1}_{0}^F$ & $\mathbf{1}_{0}^S$ & \multirow{2}{*}{A} & \multirow{2}{*}{$\mathbf{2}_{1}^F$} & \multirow{2}{*}{$\mathbf{3}_{2}^S$} & A & $\mathbf{2}_{-1}^F$ & $\mathbf{1}_{0}^S$ \\ \cline{4-6}\cline{10-12}
& & & B & $\mathbf{3}_{0}^F$ & $\mathbf{3}_{0}^S$ & & & & B & $\mathbf{2}_{-1}^F$ & $\mathbf{3}_{0}^S$ \\ \hline
\multicolumn{3}{|c||}{$(m_{\nu})_{\alpha\beta}/\langle H\rangle^3=-\frac{\mu y_{\alpha i}y'_{i\beta}}{M_{\psi_i}M_{\phi}^2}$} & \multicolumn{3}{c||}{$(m_{\nu})_{\alpha\beta}/\langle H\rangle^3=-\frac{\mu y_{\alpha i}y'_{i\beta}}{M_{\psi_i}M_{\phi}^2}$} & \multicolumn{3}{c||}{$(m_{\nu})_{\alpha\beta}/\langle H\rangle^3=-\frac{\mu y'_{\alpha i}y_{i\beta}}{M_{\psi_i}M_{\phi}^2}$} & \multicolumn{3}{c|}{$(m_{\nu})_{\alpha\beta}/\langle H\rangle^3=-\frac{\mu y'_{\alpha i}y_{i\beta}}{M_{\psi_i}M_{\phi}^2}$} \\ \hline\hline
\end{tabular}
\end{small}
\caption{\label{tab:tree_diagram}Possible topologies and diagrams for the tree level decomposition of the
dimension six effective operator $\overline{\ell_{L}}\widetilde{H}\nu_{R}\left(H^{\dagger}H\right)$.
}
\end{table}

\begin{table}[!htb]
\centering
\begin{tabular}{|m{4.5cm}<{\centering}|m{0.4cm}<{\centering}|m{0.5cm}<{\centering}|m{0.7cm}<{\centering}|m{0.7cm}<{\centering}|m{4.3cm}<{\centering}|N}\hline\hline
\multirow{-1}{*}{\includegraphics[width=0.9\linewidth]{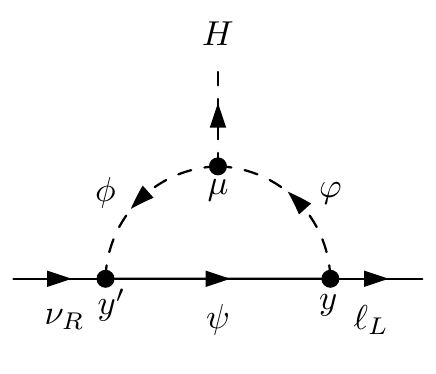}} & & $\psi$ & $\phi$ & $\varphi$ & ~~Dark Matter  \\ \cline{2-7}
& A & $\mathbf{1}_{\alpha }^F$ & $\mathbf{1}_{\alpha }^S$ & $\mathbf{2}_{\alpha+1}^S$ & $[\varphi ]_{-2}$, $[\psi ,\phi ,\varphi ]_{0}$ &  \\[0.37cm] \cline{2-7}
& B & $\mathbf{2}_{\alpha }^F$ & $\mathbf{2}_{\alpha }^S$ & $\mathbf{1}_{\alpha +1}^S$ & $[\phi ,\varphi ]_{-1}$, $[\phi ]_{1}$ & \\[0.37cm] \cline{2-7}
& C & $\mathbf{2}_{\alpha }^F$ & $\mathbf{2}_{\alpha }^S$ & $\mathbf{3}_{\alpha +1}^S$ & $[\phi ,\varphi ]_{-1}$, $[\phi, \varphi ]_{1}$ & \\[0.37cm] \cline{2-7}
E1-1 & D & $\mathbf{3}_{\alpha }^F$ & $\mathbf{3}_{\alpha }^S$ & $\mathbf{2}_{\alpha +1}^S$ & $[\phi ,\varphi ]_{-2}$, $[\psi ,\phi ,\varphi ]_{0}$ &  \\[0.37cm] \hline
\multicolumn{6}{|c|}{$(m_{\nu})_{\alpha\beta}/\langle H\rangle^3=\mu y_{\alpha i}y'_{i \beta}M_{\psi _i} I_{3}\left(M_{\psi_i },M_{\phi },M_{\varphi}\right)$}\\ \hline\hline
\multirow{-1}{*}{\includegraphics[width=0.9\linewidth]{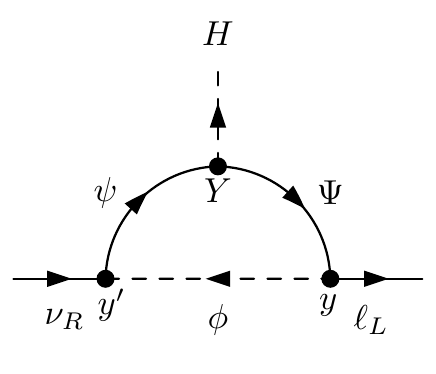}}  & & $\psi$ & $\Psi$ & $\phi$ & ~~Dark Matter  \\ \cline{2-7}
& A & $\mathbf{1}_{\alpha }^F$ & $\mathbf{2}_{\alpha-1}^F$ & $\mathbf{1}_{\alpha}^S$ & $[\psi ,\phi ]_{0}$ &  \\[0.37cm] \cline{2-7}
& B & $\mathbf{2}_{\alpha }^F$ & $\mathbf{1}_{\alpha-1}^F$ & $\mathbf{2}_{\alpha}^S$  & $[\phi ]_{-1}$, $[\Psi ,\phi ]_{1}$  & \\[0.37cm] \cline{2-7}
& C & $\mathbf{2}_{\alpha }^F$ & $\mathbf{3}_{\alpha-1}^F$ & $\mathbf{2}_{\alpha}^S$  & $[\phi ]_{-1}$, $[\Psi ,\phi ]_{1}$  & \\[0.37cm] \cline{2-7}
E1-2 & D & $\mathbf{3}_{\alpha }^F$ & $\mathbf{2}_{\alpha-1}^F$ & $\mathbf{3}_{\alpha}^S$  & $[\psi ,\phi ]_{0}$ &  \\[0.37cm] \hline
\multicolumn{6}{|c|}{$(m_{\nu})_{\alpha\beta}/\langle H\rangle^3=y_{\alpha  i} Y_{i j} y'_{j \beta} \left[M_{\Psi _i} M_{\psi_j} I_{3}\left(M_{\phi },M_{\Psi _i},M_{\psi_j}\right)+J_{3}\left(M_{\phi },M_{\Psi _i},M_{\psi_j}\right)\right]$}\\ \hline\hline
\end{tabular}
\caption{\label{tab:Ma_model}Generation of the dimension four Dirac neutrino mass term $\overline{\ell_{L}}\widetilde{H}\nu_{R}$ at one-loop~\cite{Ma:2016mwh}. The possible assignments of the mediator fields, dark matter candidates and the predictions for neutrino masses are presented.}
\end{table}

\section{\label{sec:diagrams}Systematic classifications of tree and one-loop
realizations of dimension six Dirac neutrino mass operator}

In this section, we shall show all possible decompositions of the effective
dimension six operator $\overline{\ell_{L}}\widetilde{H}\nu_{R}\left(H^{\dagger}H\right)$
for the Dirac neutrino masses at both tree level and one-loop level. For each topology and possible Lorentz structures, we shall give the quantum numbers of the mediators and the expression of the neutrino mass matrix.
In this setup, the neutrino masses are typically estimated as
\begin{equation}
m_{\nu}\sim C\left(\frac{1}{16\pi^2}\right)^{j}\frac{v^3}{\Lambda^2}\,,
\end{equation}
where $j=0$ for tree level diagrams and $j=1$ for one-loop diagrams, $v$ denotes the vacuum expectation value of the SM Higgs field, $\Lambda$ is the new physics scale. All coupling constants
%, and for some models also certain mass-scale ratios,
are absorbed in the dimensionless coefficient $C$. To accommodate the neutrino mass scale $\mathcal{O}(1)$ eV, the new physics scale should be approximately
\begin{equation}
\Lambda\sim6\times10^{3}C^{1/2}\,\text{TeV}\,,
\end{equation}
for one-loop realizations. In a concrete model, some of the constants are expected to be less than one, which could bring the new physics scale to lower values.

\subsection{\label{subsec:tree}Tree level}

There are only two possible topologies named as F1 and F2 at tree level, as shown in table~\ref{tab:tree_diagram}. The
new fields are denoted by the notation $X^{\mathcal{L}}_Y$, where $X$
corresponds to the $SU(2)$ representation under which the field transforms,
$\mathcal{L}$ refers to the Lorentz nature, $S$ for scalar and $F$ for
fermion, $Y$ is the hypercharge fulfilling $Y=2(Q - T_3)$, where $Q$ is the electric charge and $T_3$ is its isospin. In this work, we
shall consider the case that the new fields are electroweak singlet
$X=\mathbf{1}$, doublet $X=\mathbf{2}$ or triplet $X=\mathbf{3}$. The
results for larger representations can be easily obtained. The new fields
are assumed to be either scalars or fermions, the fermions should be vector-like to ensure anomaly cancellation. The diagrams with scalar
or vector bosons are equivalent, and the resulting neutrino masses for the
diagrams with vectors can be straightforwardly obtained from those of the
diagrams with scalars. Furthermore, vector bosons are generally the gauge bosons of a certain gauge symmetry, their mass are generated via the spontaneous breaking of that symmetry. As a result, the scalar sector of these models should be discussed carefully as well, the corresponding analysis is highly model dependent. Therefore we shall focus on the scalar and fermion mediated models in the present work.

We present our results in table~\ref{tab:tree_diagram}, where the transformation rules of the messenger fields are determined from the invariance of each interaction vertex under the SM gauge group. For the topology F1, only one diagram is allowed and a new scalar field transforming as $\mathbf{2}^{S}_1$ should be introduced. For the second topology F2, there
are altogether nine possible realizations leading to the effective operators
$\overline{\ell_{L}}\widetilde{H}\nu_{R}\left(H^{\dagger}H\right)$.
One needs to introduce two mediators, either of which can be scalar or
fermion field. We see that the mediators can take two different
sets of quantum numbers for certain cases (e.g. F2-1-2).

\subsection{\label{sebsec:one_loop}One-loop level}

\begin{figure}[!htb]
\centering
\includegraphics[width=0.99\linewidth]{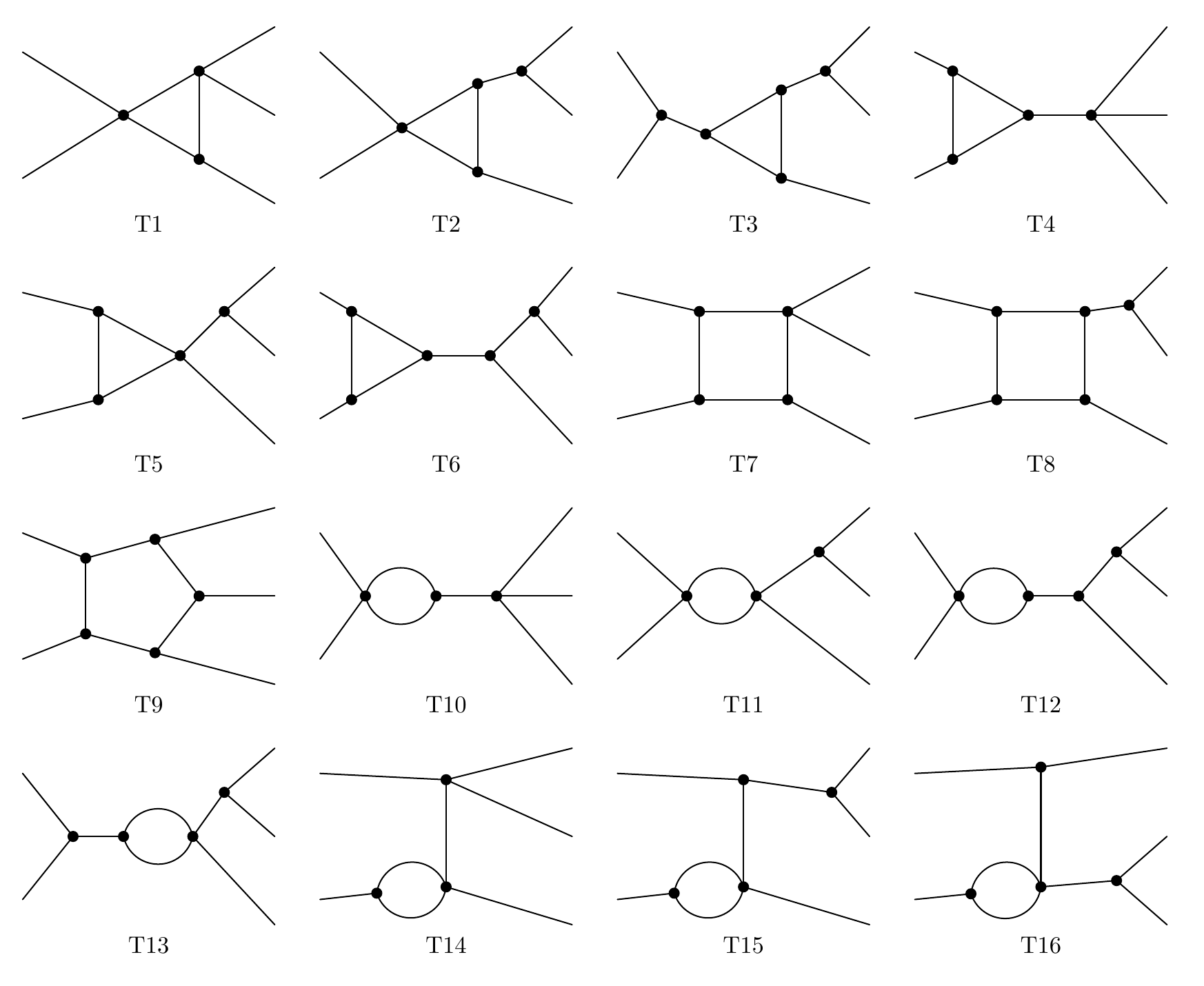}
\caption{\label{fig:topfig}Topologies of one-loop diagrams with five external legs.}
\end{figure}

A systematical analysis of how to generate the renormalizable Dirac neutrino mass operator  $\overline{\ell_{L}}\widetilde{H}\nu_{R}$
at one-loop level has been performed in Ref.~\cite{Ma:2016mwh,Wang:2016lve}. It was shown that there are only two possible topologies, which are displayed in table~\ref{tab:Ma_model} and would be called Ma diagrams hereafter. The SM gauge quantum numbers of the mediator fields, the possible dark matter candidates and the expressions of the neutrino masses are presented as well. In this section, we shall follow the diagram-based approach of Ref.~\cite{Bonnet:2012kz,Sierra:2014rxa} to find out all possible one-loop realizations for dimension six Dirac neutrino mass operator. Firstly we use the program {\tt FeynArts}~\cite{Hahn:2000kx} to construct the one-loop topologies with five external legs, the self-energy and tadpole diagrams are excluded. We require that the underlying theory is renormalizable such that all the vertices should contain only three or four legs. It turns out that there are totally 16 distinct topologies, as shown in figure~\ref{fig:topfig}. Subsequently we insert scalars and fermions into each topology. The topologies T1, T10 and T14 can be discarded because non-renormalizable vertices are needed in these diagrams. There are usually more than one possibilities of assigning the five external legs to the lepton doublet $\ell_L$, the right-handed neutrino singlet $\nu_R$, two Higgs doublets $H$ and the Higgs conjugate $H^{\dagger}$, therefore a given topology can give rise to several Feynman diagrams. All possible Feynman diagrams are generated with the help of \texttt{FeynArts}. The one-loop Feynman diagrams are named as Ta-b-c, where ``a'' represents the
topology, ``b'' denotes the different choices of the fermion lines
in a given topology, and ``c'' refers to the assignments of the external fields, i.e., the flow directions of $l_{L}$, $\nu_{R}$ and $H$.

\begin{figure}[t!]
\centering
\includegraphics[width=0.8\linewidth]{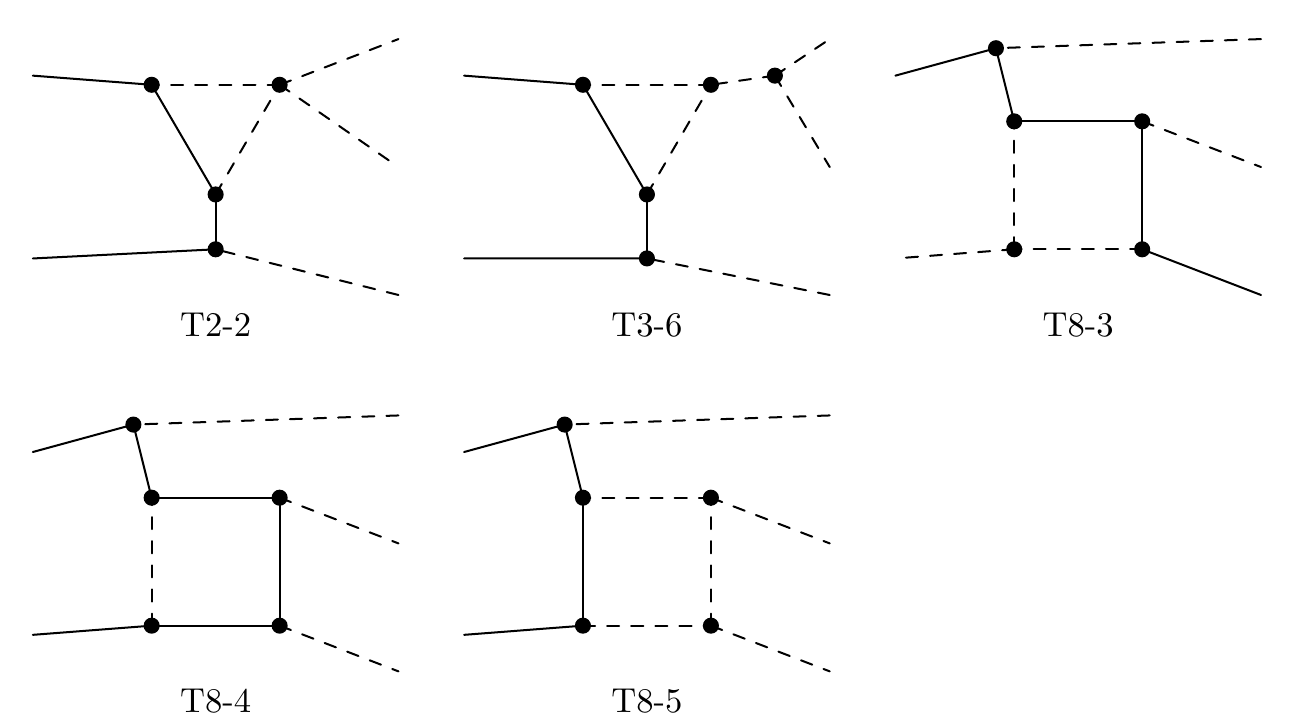}
\caption{\label{fig:top_FS_insertion_nonreducible} Finite one-loop diagrams for the Dirac neutrino mass operator $\overline{\ell_{L}}\widetilde{H}\nu_{R}\left(H^{\dagger}H\right)$. Both the tree level realizations in table~\ref{tab:tree_diagram} and the one-loop diagrams in table~\ref{tab:Ma_model} can be absent for certain quantum numbers of the messengers in these cases. The dashed lines denote scalars, and the solid lines denote fermions.}
\end{figure}

We can divide all the Feynman diagrams into two categories: the diagrams which lead to finite loop integrals and the ones which involve infinite loop integrals. We collect the divergent diagrams in figure~\ref{fig:divergent}, the corresponding divergence can be absorbed by the counter terms of the tree level realizations of the dimension six operator. For the remaining diagrams with finite loop integrals, we can identify the diagrams for which both the tree level realizations and Ma diagrams can be avoided for certain quantum numbers of the mediators. There are only five such diagrams T2-2, T3-6, T8-3, T8-4 and T8-5, as shown in figure~\ref{fig:top_FS_insertion_nonreducible}. For the sake of completeness, figure~\ref{fig:top_FS_insertion_reducible} shows other finite diagrams for which either the tree level contributions or the Ma diagrams can not be forbidden if no additional symmetry is introduced.
Furthermore, we consider all possible assignments for the fermion and scalar internal lines as well as the external lines of each diagram in figure~\ref{fig:top_FS_insertion_nonreducible}. Eventually we find only 13 irreducible Feynman diagrams which are suitable to generate the small Dirac
neutrino masses at one-loop level. The quantum numbers of the mediator fields and the predictions for the neutrino masses are summarized in
tables~\ref{tab:figT2}\,--\,\ref{tab:figT8}. Here we have focused on the $SU(2)$
singlets, doublets and triplets of scalars or fermions. Results for larger representations can be easily obtained. Notice that we don't give explicitly the color quantum numbers. Since both lepton and Higgs doublets are color singlets, the color charges of the mediator particles can be straightforwardly fixed from $SU(3)$ multiplication rules $\mathbf{1}\otimes\mathbf{1}=\mathbf{1}$, $\mathbf{3}\otimes\bar{\mathbf{3}}=\mathbf{1}\oplus\mathbf{8}$ and $\mathbf{8}\otimes\mathbf{8}=\mathbf{1}\oplus\mathbf{8}\oplus\overline{\mathbf{10}}\oplus\mathbf{8}\oplus\mathbf{10\oplus\mathbf{27}}$ etc. On the other hand, we would like to discuss possible dark
matter candidates in our models, it is well-known that dark matter particle is color neutral, therefore all states are assumed to be color singlets in this work.

From tables~\ref{tab:figT2}\,--\,\ref{tab:figT8}, we see that the hypercharges of the extra fields are fixed up to a free real parameter $\alpha$, and $\alpha$ should be an even or odd integer to avoid fractionally charged particles. It is remarkable that both tree level diagrams shown in table~\ref{tab:tree_diagram} and the Ma diagrams in table~\ref{tab:Ma_model} can be absent for certain values of $\alpha$, such that the neutrino masses dominantly arise from the effective dimension six operator at one-loop level. We list the excluded values of $\alpha$ from the appearance of lower order tree diagrams and Ma diagrams in these tables, where $\mathbb{U}$ denotes the universal set (i.e., the set of all possible values). If the excluded values of $\alpha$ constitute the universal set $\mathbb{U}$, the lower order tree or Ma diagrams can not be avoided without additional symmetry. For completeness we present all possible assignments of the mediators for a given diagram in tables~\ref{tab:figT2}\,--\,\ref{tab:figT8}, the models accompanied by lower order neutrino masses are not neglected.

The tables~\ref{tab:figT2}\,--\,\ref{tab:figT8} are useful to read off the potential interesting radiative neutrino mass models. In the following, we take T2-2-1-A as a concrete example. The new fields of this model include two SM singlets $\psi\sim\mathbf{1}^{F}_{\alpha}$ and $\phi\sim\mathbf{1}^{S}_{\alpha}$, and two triplets $\Psi\sim\mathbf{3}^{F}_{-2}$ and $\varphi\sim\mathbf{3}^{S}_{\alpha+2}$. The relevant Feynman diagram can be generated from the following Lagrangian invariant under the SM gauge symmetry,
\begin{equation}
\begin{split}
\mathscr{L}_{\text{T2-2-1-A}}= &\left[(y_{L})_{\alpha i}\overline{\ell_{L}}_\alpha(\tau^a\Psi_i^a)H+y_{ij}\overline{\Psi_i^a}\psi_j\varphi^{*a}+(y_R)_{j\beta}\overline{\psi_j}\nu_{R\beta}\phi+\lambda H^\dagger (\tau^a\varphi^a)\widetilde{H}\phi^*+\text{H.c.}\right]\\
&-M_\psi \overline{\psi}\psi-M_\Psi \overline{\Psi^a}\Psi^a-M_\phi^2\phi^\dagger\phi-M_\varphi^2\varphi^{\dagger a}\varphi^a\,,
\end{split}
\end{equation}
where $\alpha, \beta, i, j=1, 2, 3$ are flavor indices, and we assume there are three generations of new fermions $\psi$ and $\Psi$. The index $a$ is for the adjoint representation~($a=1,2,3$), $y_L, y_R$ and $y$ denote the Yukawa couplings at the new interaction vertices, and $\lambda$ is the coupling constant among the Higgs field $H$ and the new scalars $\phi$ and $\varphi$.

As regards the neutrino masses, we first calculate the one-loop Feynman diagram of T2-2-1 before the electroweak symmetry breaking to obtain the effective interaction vertex with five external legs, the predictions for the neutrino masses follows immediately after the Higgs field $H$ acquires vacuum expectation value. Since the momentum of the external lines ($\ell_L$, $\nu_{R}$ and $H$) are irrelevant to the neutrino masses, we can set them to be zero, and then the loop integral would be greatly simplified. Eventually we find the expression for the neutrino mass matrix is given as
\begin{equation}
(m_{\nu})_{\alpha\beta}/\langle H\rangle^3=-\frac{M_{\psi _j}}{M_{\Psi _i}}\lambda
{(y_L)}_{\alpha  i} y_{i j} {(y_R)}_{j \beta }I_{3}\left(M_{\phi
},M_{\varphi },M_{\psi _j}\right)\,,
\end{equation}
where the function $I_3$ is given by Eq.~\eqref{eq:I3}. Nine distinct loop integrals $I_{2,3,4,5}, J_{3,4,5}$ and $K_{4,5}$ are needed when evaluating the one-loop Feynman diagrams in this work, their explicit forms are collected in Appendix~\ref{sec:app_mass}. For all other possible one-loop models, we can follow the same procedure to write down the Lagrange of the model and extract the resulting predictions for neutrino masses.

\section{\label{sec:forbidtree}Possible schemes to forbid lower order contributions}

If the effective dimension six operator $\overline{\ell_{L}}\widetilde{H}\nu_{R}\left(H^{\dagger}H\right)$ arising from one-loop Feynman diagram is the leading order contribution to the neutrino masses, it is necessary to ensure that the lower order contributions are absent. For instance, both the tree level realizations in table~\ref{tab:tree_diagram} and the Ma diagrams in table~\ref{tab:Ma_model} should be forbidden. Moreover, one should prevent the tree-level term generated by the renormalizable Yukawa interaction $\overline{\ell_L}\widetilde{H}\nu_{R}$
in radiative Dirac neutrino mass models~\cite{Cai:2017jrq,Ma:2016mwh}. In general a symmetry is needed to avoid the appearance of the above mentioned lower order contributions. In the following, we shall present two possible strategies to deal with this issue: the first scheme is the soft breaking abelian symmetry~\cite{Ma:2016mwh} and the second one is the finite non-abelian symmetry.

\begin{figure}[h]
\centering
\includegraphics[width=0.99\textwidth]{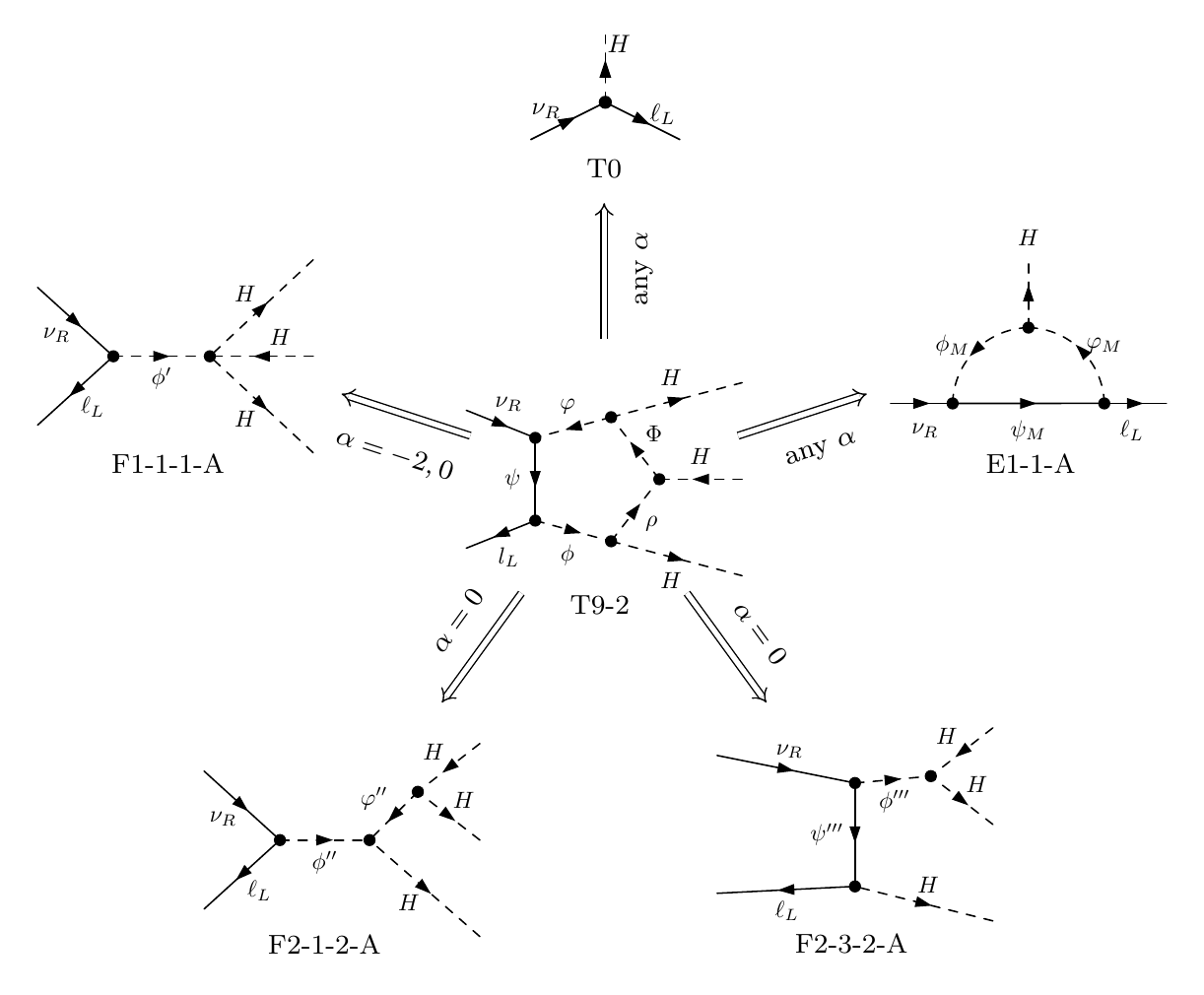}
\caption{Five possible lower order diagrams which are present in the model based on the pentagon diagram T9-2.}
\label{fig:starfish}
\end{figure}

In order to illustrate this idea, we take the diagram T9-2 as an example, which is shown in figure~\ref{fig:starfish}. We see that five messenger fields are needed and they transform under the SM gauge symmetry as $\psi\sim \mathbf{1}_{\alpha}^F$, $\varphi, \rho\sim \mathbf{1}_{\alpha}^S$
and $\phi,\Phi\sim \mathbf{2}_{\alpha+1}^S$. Comparing with the field content of the Ma model E1-1-A shown in table~\ref{tab:Ma_model}, we can identify $\psi_M\sim \psi$, $\phi_M\sim\varphi (\rm{or}~\rho)$ and $\varphi_M\sim\phi(\rm{or}~\Phi)$ where the subscript ``$M$'' indicates the fields in the Ma model. This means that the diagram E1-1-A is always present for any value of $\alpha$ in the T9-2 model. As a consequence, the diagram T9-2 only gives a subleading contribution to the neutrino masses with respect to E1-1-A. Tree level diagrams can also appear in T9-2 model. For example, when $\alpha=-2$, $\phi^{*},\Phi^{*}\sim \mathbf{2}_{1}^S$ can be used as the mediator of F1-1-1-A. Similarly we see that the tree level diagrams F1-1-1-A, F2-1-2-A, and F2-3-2-A can not be avoided for the case of $\alpha=0$.

If we require that the neutrino masses dominantly arise from the diagram T9-2, the above mentioned five diagrams T0, E1-1-A, F1-1-1-A, F2-1-2-A, and F2-3-2-A should be absent. These five lower order contributions can really be avoided by introducing a soft breaking $Z_3$ symmetry. The lepton doublet $\ell_{L}$, right-handed neutrino $\nu_{R}$ and Higgs field $H$ are assigned to transform as $\ell_{L}\rightarrow\omega\ell_{L}$, $\nu_{R}\rightarrow\omega\nu_{R}$ and $H\rightarrow\omega H$, where $\omega=e^{i2\pi/3}$ is the cube root of unit. Thus the renormalizable coupling $\overline{\ell_{L}}\widetilde{H}\nu_{R}$ as well as the tree level
diagram T0 would be excluded. Because the combination $(H^\dagger H)$ is a singlet under $Z_3$, all the effective Dirac neutrino mass operators shown in Eq.~\eqref{eq:lag_d4} would be forbidden as well if $Z_3$ is an exact symmetry. This implies that the $Z_3$ symmetry should be broken. For simplicity, we assume that $Z_3$ is softly broken at only one interaction vertex of the T9-2 diagram. Then there are five types of $Z_3$ charge assignments for the mediator fields, as listed in table~\ref{tab:qnumber}. The fields involved in the soft breaking vertex are indicated with the symbol ``$\blacktriangle$'' as the superscript of the $Z_3$ charges. We see that all tree level diagrams and Ma models are forbidden for the assignment $\psi, \Phi\sim 1$, $\rho\sim\omega$ and $\phi, \varphi\sim \omega^2$, unless more than one $Z_3$ soft breaking interactions are introduced.

\begin{table}[hptb!]
  \centering
  \begin{tabular}{|c|c|c|c||c|c|c|c|c||c|c|c|c|}
    \hline\hline
   & \multicolumn{3}{c||}{T0} &  \multicolumn{5}{c||}{T9-2} &
   \multicolumn{3}{c|}{E1-1-A} & \\ \cline{1-12}
   \text{Fields} & $\ell_L$ & $\nu_R$ & $H$ & $\psi$ & $\phi$ & $\varphi$ &
   $\Phi$ & $\rho$ & $\psi_M$ & $\phi_M$ & $\varphi_M$ & Viable \\
   \cline{1-12}
    \text{Gauge Sym.} & $\mathbf{2}_{-1}^F$ & $\mathbf{1}_{0}^F$ & $\mathbf{2}_{1}^S$ & $\mathbf{1}_{\alpha }^F$
    & $\mathbf{2}_{\alpha +1}^S$ & $\mathbf{1}_{\alpha }^S$ & $\mathbf{2}_{\alpha +1}^S$ & $\mathbf{1}_{\alpha
    }^S$ & $\mathbf{1}_{\alpha }^F$ & $\mathbf{1}_{\alpha }^S$ & $\mathbf{2}_{\alpha +1}^S$ &  \\
    \hline
    \multirow{5}{*}{$Z_3$~\text{Sym.}} & \multirow{5}{*}{$\omega$} &
    \multirow{5}{*}{$\omega$} & \multirow{5}{*}{$\omega$} & $1_\blacktriangle$ &
    $1_\blacktriangle$ & $\omega^2$ & $1$ & $\omega^2$ & $1_\blacktriangle$ & $\omega^2$ &
    $1_\blacktriangle$ & \ding{55}\\ \cline{5-13}
    &  &  &  & $1$ & $\omega^2_{\blacktriangle}$ & $\omega^2$ & $1$ &
    $\omega^2_{\blacktriangle}$ & $1$ & $\omega^2_{\blacktriangle}$ & $\omega^2_{\blacktriangle}$ &
    \ding{55}\\ \cline{5-13}
    &  &  &  & $1$ & $\omega^{2}$ & $\omega^2$ & $1_\blacktriangle$ &
    $\omega_{\blacktriangle}$ & $1$ & $\omega_{\blacktriangle}$ & $1_{\blacktriangle}$ & \ding{51}\\
    \cline{5-13}
    &  &  &  & $1$ & $\omega^{2}$ & $\omega^2_{\blacktriangle}$ & $\omega^2_{\blacktriangle}$
    & $\omega$ & $1$ & $\omega^2_{\blacktriangle}$ & $\omega^2_{\blacktriangle}$ & \ding{55}
    \\ \cline{5-13}
     &  &  &  & $1_\blacktriangle$ & $\omega^{2}$ & $\omega_{\blacktriangle}$ & $\omega^2$ &
     $\omega$ & $1_\blacktriangle$ & $\omega_{\blacktriangle}$ & $\omega^2$ & \ding{55}\\
     \hline\hline
  \end{tabular}
\caption{\label{tab:qnumber}Forbidding lower order contributions by using soft breaking $Z_3$, where $\omega=e^{i2\pi/3}$. The subscript ``$\blacktriangle$'' denotes that the corresponding fields are involved in the $Z_3$ soft breaking vertex.  }
\end{table}

\begin{figure}[h]
\centering
\includegraphics[width=0.95\linewidth]{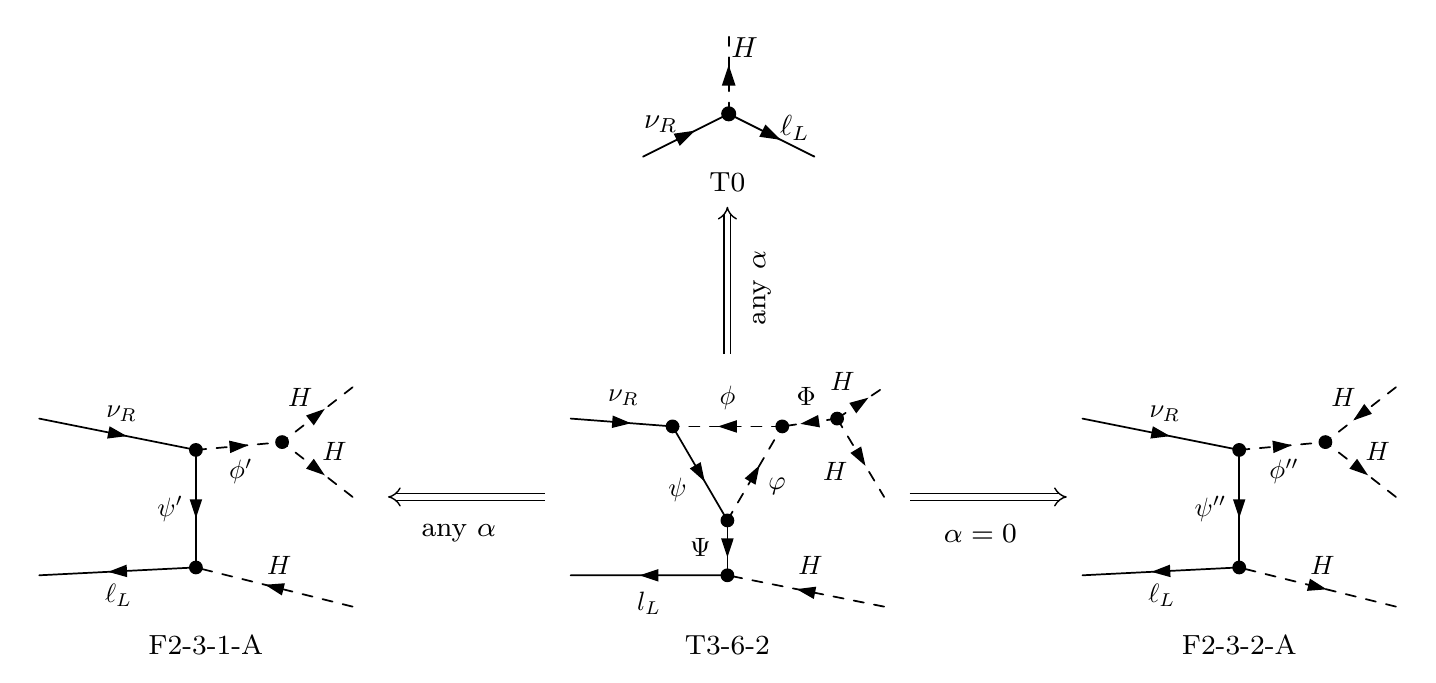}
\caption{Three possible lower order diagrams allowed in the T3-6-2 model.}
\label{fig:t361_dec}
\end{figure}

Besides soft breaking abelian symmetry, finite non-abelian symmetry can also help to forbid the undesired diagrams which give a larger contribution to neutrino masses. In order to show this scheme can really work, we shall apply the $S_4$ symmetry group to the T3-6-2 model. As displayed in figure~\ref{fig:t361_dec}, the T3-6-2 model contains five mediators $\psi\sim\mathbf{1}^{F}_{\alpha}$, $\Psi\sim\mathbf{3}^{F}_{-2}$, $\phi\sim\mathbf{1}^{S}_{\alpha}$, $\varphi\sim\mathbf{3}^{S}_{\alpha+2}$ and $\Phi\sim\mathbf{3}^{S}_{-2}$ such that the tree diagrams T0 and F2-3-1-A always exist for any value of $\alpha$, and a third tree diagram F2-3-2-A appears for $\alpha=0$.

The $S_4$ group has been extensively used to predict lepton flavor mixing angles and CP violation phases~\cite{Ma:2005pd,Ding:2009iy,Hagedorn:2010th,Ding:2013hpa,Ishimori:2010au}. $S_4$ is the permutation group of four distinct objects, and it is isomorphic to the symmetry group of a regular octahedron. The $S_4$ group has five irreducible representations:  two singlets $\mathbf{1}$ and $\mathbf{1}^{\prime}$, one doublet $\mathbf{2}$ and two triplets $\mathbf{3}$ and $\mathbf{3}^{\prime}$~\cite{Ma:2005pd,Ding:2009iy,Hagedorn:2010th,Ding:2013hpa,Ishimori:2010au}. The Kronecker products between these irreducible
representations are as follows,
\begin{eqnarray}
\nonumber&& \mathbf{1}\otimes R=R\otimes\mathbf{1}=R,\quad
\mathbf{1^{\prime}}\otimes\mathbf{1^{\prime}}=\mathbf{1},\quad
\mathbf{1^{\prime}}\otimes\mathbf{2}=\mathbf{2},\quad
\mathbf{1^{\prime}}\otimes\mathbf{3}=\mathbf{3^{\prime}},\quad
\mathbf{1^{\prime}}\otimes\mathbf{3^{\prime}}=\mathbf{3},\\
\nonumber&&
\mathbf{2}\otimes\mathbf{2}=\mathbf{1}\oplus\mathbf{1^{\prime}}\oplus\mathbf{2},\quad
\mathbf{2}\otimes\mathbf{3}=\mathbf{2}\otimes\mathbf{3^{\prime}}=\mathbf{3}\oplus\mathbf{3^{\prime}},\\
&&\mathbf{3}\otimes\mathbf{3}=\mathbf{3^{\prime}}\otimes\mathbf{3^{\prime}}=\mathbf{1}\oplus\mathbf{2}\oplus\mathbf{3}\oplus\mathbf{3^{\prime}},\quad
\mathbf{3}\otimes\mathbf{3^{\prime}}=\mathbf{1^{\prime}}\oplus\mathbf{2}\oplus\mathbf{3}\oplus\mathbf{3^{\prime}}\,,
\end{eqnarray}
where $R$ stands for any representation. We assign the three generations of left-handed lepton doublets $\ell_{L}$ and right-handed neutrinos $\nu_{R}$ to two triplets $\mathbf{3}$ and $\mathbf{3}^{\prime}$ respectively, and the Higgs field $H$ transforms as $\mathbf{1}$. The transformation properties of the mediators under $S_4$ are shown in table~\ref{tab:qnumber_s4_v2}. Obviously $S_4$ prevents the direct Yukawa coupling T0. For the tree diagram F2-3-1-A, the field $\psi^{\prime}$ is uniquely identified with $\Psi$, while $\phi^{\prime}$ can be identified with $\Phi^{*}$ irrespective of $\alpha$ value or $\varphi$ when $\alpha=0$. For the former case, the vertex involving $\psi'$, $\phi'$ and $\nu_R$ would break the $S_4$ symmetry, and for the latter case, the vertex involving $\phi'$ and $H$ would break the $S_4$ symmetry. As a result, the tree diagram F2-3-1-A is forbidden. Regarding the decomposition into F2-3-2-A in case of $\alpha=0$, $\psi''$
and $\phi''$ should transform in the same way as $\psi$ and $\phi$ respectively. Then both interaction vertices $H^{\dagger}H\phi''$ and $\bar{\ell}_{L}\widetilde{H}\psi''$ in the diagram F2-3-2-A would be forbidden by $S_4$ although they are invariant under the SM gauge symmetry. Therefore the Feynman diagram T3-6-2 would give the leading contribution to neutrino masses after the $S_4$ symmetry is imposed .

\begin{table}[hptb!]
  \centering
  \begin{tabular}{|c|c|c|c||c|c|c|c|c||c|c||c|c|c|}
    \hline\hline
    & \multicolumn{3}{c||}{T0} &  \multicolumn{5}{c||}{T3-6-2} & \multicolumn{2}{c||}{F2-3-1-A} & \multicolumn{2}{c|}{F2-3-2-A} \\ \hline
   \text{Fields} & $\ell_L$ & $\nu_R$ & $H$ & $\psi$ & $\Psi$ & $\phi$ & $\varphi$ & $\Phi$ & $\psi'$ & $\phi'$ & $\psi''$ & $\phi''$ \\ \hline
   \text{Gauge Sym.} & $\mathbf{2}_{-1}^F$ & $\mathbf{1}_{0}^F$ & $\mathbf{2}_{1}^S$ & $\mathbf{1}_{\alpha}^F$ & $\mathbf{3}_{-2}^F$ & $\mathbf{1}_{\alpha }^S$ & $\mathbf{3}_{\alpha +2}^S$ & $\mathbf{3}_{-2}^S$ & $\mathbf{3}_{-2}^F$ & $\mathbf{3}_{2}^S$ & $\mathbf{1}_{0}^F$ & $\mathbf{1}_{0}^S$   \\ \hline
   \text{$S_4$ Sym.} & $\mathbf{3}$ & $\mathbf{3^\prime}$ & $\mathbf{1}$ & $\mathbf{3}^{\prime}$ & $\mathbf{3}$ & $\mathbf{3}$ & $\mathbf{3}$ & $\mathbf{1}$ & $\mathbf{3}$ & $\mathbf{1}$ (or $\mathbf{3}$) & $\mathbf{3}^{\prime}$ & $\mathbf{3}$  \\ \hline\hline
\end{tabular}
\caption{\label{tab:qnumber_s4_v2} Forbidding lower order contributions by using non-abelian symmetry $S_4$. Here we list the transformation properties of the mediator fields under $S_4$.}
\end{table}

\section{\label{sec:darkmatter}Dark matter candidates}

In above neutrino models, the new messenger fields could possibly be identified as dark matter particles such that neutrino masses and dark matter can be accounted for in a single model. This attractive idea is firstly proposed in Ref.~\cite{Ma:2006km}. A comprehensive analysis of one-loop Majorana neutrino mass models with viable dark matter candidates has been performed in~\cite{Restrepo:2013aga}. Some models connecting Dirac neutrino masses to dark matter have been proposed~\cite{Gu:2007gy,
Farzan:2012sa,Bonilla:2016diq,Kanemura:2017haa}. In the following, we shall investigate under what conditions the above one-loop model for Dirac neutrino masses can accommodate the dark matter.

It is universally recognized that the dark matter candidate should be stable, colorless and not charged with electricity. In order to guarantee the stability of dark matter, generally an extra symmetry is imposed to distinguish the dark matter particle from the SM ones. The discrete $Z_2$ symmetry is widely used in the radiative neutrino masses models which can account for dark matter. We shall assume that any of the particles in the loop transforms as odd under the $Z_2$ while the SM particles and other mediator fields are even under the $Z_2$. As a consequence, only the fields mediating the loop could be dark matter candidates. For example, for the model T2-2-1, the $Z_2$ parity of the internal messengers $\psi$, $\phi$ and $\varphi$ are odd, yet the $Z_2$ parity of $\Psi$ as well as SM fields is even. Notice that this dark matter $Z_2$ parity can also help to forbid the contributions from the tree level diagrams and Ma diagrams.

The existence of electrically neutral particle in the spectrum is an important restriction. Solving the condition $Q=T_3+Y/2=0$, we arrive at $Y=-2T_3$. That is to say only the component satisfying $Y=-2T_3$ inside a multiplet could be the dark matter candidate. As regards the concerned one-loop models for Dirac neutrino masses, the condition $Y=-2T_3$ would fixes the parameter $\alpha$ to be few discrete values. This point can be clearly seen from the tables~\ref{tab:figT2}\,--\,\ref{tab:figT8} in Appendix~\ref{sec:app_diagram}. Each such value of $\alpha$ would define a different model with a given field content.

Dark matter direct detection plays a significant role in our analysis, because the direct coupling between the dark particle and $Z$ boson is proportional to the hypercharge $Y$. Consequently the dark matter particle can elastically scatter with nuclei through $Z$ boson exchange diagram. The resulting spin-independent cross section would be several order of magnitude larger than current bounds from LUX~\cite{Akerib:2016vxi} and XENON1T~\cite{Aprile:2017iyp}, if the hypercharge $Y$ is nonzero. Taking into account the electrically neutral condition further, we obtain that the dark matter particle must satisfy $Y=T_3=0$. This requirement excludes multiplets with even number of fields (i.e., doublets or quartets etc.), and the dark matter candidates can only be singlet or triplet scalars and fermions with zero hypercharge. An exception is the case of the scalar doublet with hypercharge $Y=\pm1$. The scalar potential of the model allows to have a mass splitting between the scalar and pseudo-scalar neutral states. The lighter one can be the dark matter candidate and its scattering off a nucleus via $Z$-mediated diagrams can be kinematically forbidden in the case that its kinetic energy is less than the mass difference of the scalar and pseudo-scalar fields~\cite{Farzan:2012ev,LopezHonorez:2006gr,Honorez:2010re,Goudelis:2013uca}.

Now let us take the model T2-2-1-A as an example to illustrate whether this model can account for the dark matter and under what conditions. This model needs four messenger fields: two $SU(2)$ singlets $\psi\sim \mathbf{1}_{\alpha }^F$, $\phi\sim \mathbf{1}_{\alpha }^S$ and two triplets $\Psi\sim\mathbf{3}_{-2}^F$, $\varphi\sim \mathbf{3}_{\alpha+2}^S$. We find that three different values of $\alpha$ can lead to a neutral particle in the spectrum.

\begin{itemize}
\item $\alpha=0$: \quad $\psi_0^0$, \quad
    $\Psi_{-2}=(\Psi^0,\Psi^-,\Psi^{--})$, \quad $\phi_0^0$,\quad
    $\varphi_2=(\varphi^{++},\varphi^{+},\varphi^{0})$

In this case the dark matter candidate is a mixture of the neutral components from the singlet and triplet scalars or a mixed singlet-triplet fermion. Note that the spectrum contains a doubly charged fermion $\Psi^{--}$, and it leads to background-free signals which can be searched for at the LHC. Both fermion fields $\psi_0^0$ and $\Psi_{-2}$ should be vector-like to ensure the anomaly cancellation.

\item $\alpha=-2$: \quad $\psi_{-2}^-$, \quad
    $\Psi_{-2}=(\Psi^0,\Psi^-,\Psi^{--})$, \quad $\phi_{-2}^-$,\quad
    $\varphi_0=(\varphi^{+},\varphi^{0},\varphi^{-})$

This model allows for a triplet scalar dark matter, and
fermion dark matter is excluded by direct detection bounds. The anomaly cancellation requires that both fermions $\psi_{-2}^-$ and  $\Psi_{-2}$ must be vector-like.

\item $\alpha=-4$: \quad $\psi_{-4}^-$, \quad
    $\Psi_{-2}=(\Psi^0,\Psi^-,\Psi^{--})$, \quad $\phi_{-4}^-$,\quad
    $\varphi_{-2}=(\varphi^{0},\varphi^{-},\varphi^{--})$

The neutral particles belong to scalar and fermion triplets with non-zero hypercharge such that the condition $Y=T_3=0$ can not be fulfilled. Therefore this possibility is not consistent with dark matter.

\end{itemize}

In the same fashion the possible dark matter candidates can be studied for the other viable radiative neutrino mass models, the corresponding results are listed in the last column of the tables~\ref{tab:figT2}\,--\,\ref{tab:figT8}. We see that in most cases the values of the parameter $\alpha$ consistent with dark matter are excluded by the requirement that lower order contributions (tree level diagrams and Ma diagrams) should disappear in a given model. However, if we take into account the dark matter $Z_2$ symmetry, almost all values of $\alpha$ excluded by the lower order neutrino masses are admissible, and these numbers are shadowed in grey. As a result, neutrino masses and dark matter can be simultaneously accommodated in these models. We would like to point out that the T2-2-1-A model with $\alpha=-2$ is quite interesting, the dark matter is the neutral component of the triplet scalar $\varphi\sim\mathbf{3}^{S}_{0}$, and the lower order contributions to neutrino masses don't exist even if without $Z_2$ symmetry. Notice that the number of additional fields can be reduced for the values of $\alpha$ consistent with dark matter. For example, for the model T2-2-1-B with $\alpha=-1$, the dark matter particles can be the mixture of the neutral components of $\phi$ and $\varphi$. Moreover, since both $\varphi$ and $\phi^*$ transform as $\sim \mathbf{2}_{1}^S$, and they are odd under $Z_2$, consequently they can be identified as the same field. If a pair of mediators up to charge conjugation transform in the same way under the SM gauge symmetry and carry the same dark matter $Z_2$ parity, the symbol ``$\diamond$" would be denoted on the superscript and the corresponding values of $\alpha$ is labelled as the subscript outside the square bracket. The smaller the number of messengers the simpler the model is. We see that at least three additional multiplets are needed in the present models for neutrino masses and dark matter, and there are only 12 models which contain three new fields, as summarized in table~\ref{tab:classify_model_min}. These models are good starting points to further discuss the neutrino, dark matter and collider phenomenology in detail.

\begin{table}[hptb!]
  \centering
\begin{tabular}{|c|c|c|c|c|}
\hline\hline
\multirow{2}{*}{Model} & \multirow{2}{*}{$\alpha$} & \multicolumn{2}{c|}{Dark matter} & \multirow{2}{*}{Additional fields  }   \\ \cline{3-4}
&  & Fermionic & Scalar &  \\
\hline
T2-2-1-C & $-1$ & --- & $\phi (\varphi )\sim \mathbf{2}_ {-1}^S$ & $\psi \sim \mathbf{2}_ {-1}^F,~\Psi \sim \mathbf{3}_ {-2}^F$ \\ \hline
T2-2-2-A & $0$ & $\psi \sim \mathbf{1}_ 0^F$ & $\phi (\varphi )\sim \mathbf{1}_ 0^S$ & $\Psi \sim \mathbf{1}_ 0^F$ \\ \hline
\multirow{2}{*}{T2-2-2-C} & $-1$ & --- & $\phi (\varphi )\sim \mathbf{2}_ {-1}^S$  & $\psi \sim \mathbf{2}_ {-1}^F,~\Psi \sim \mathbf{1}_ 0^F$ \\ \cline{2-5}
 & $1$ & --- & $\phi (\varphi )\sim \mathbf{2}_ 1^S$  & $\psi \sim \mathbf{2}_ 1^F,~\Psi \sim \mathbf{1}_ 0^F$ \\ \hline
\multirow{2}{*}{T2-2-2-D} & $-1$ & --- & $\phi (\varphi )\sim \mathbf{2}_ {-1}^S$ & $\psi \sim \mathbf{2}_ {-1}^F,~\Psi \sim \mathbf{3}_ 0^F$ \\ \cline{2-5}
 & $1$ & --- & $\phi (\varphi )\sim \mathbf{2}_ 1^S$  & $\psi \sim \mathbf{2}_ 1^F,~\Psi \sim \mathbf{3}_ 0^F$ \\ \hline
T2-2-2-E & $0$ & $\psi \sim \mathbf{3}_ 0^F$ & $\phi (\varphi )\sim \mathbf{3}_ 0^S$  & $\Psi \sim \mathbf{1}_ 0^F$ \\ \hline
T2-2-2-G & $0$ & $\psi \sim \mathbf{3}_ 0^F$ & $\phi (\varphi )\sim \mathbf{3}_ 0^S$  & $\Psi \sim \mathbf{3}_ 0^F$ \\ \hline
T2-2-3-A & $-1$ & $\psi \sim \mathbf{1}_ 0^F$ & $\phi (\varphi )\sim \mathbf{2}_ 1^S$  & $\Psi \sim \mathbf{2}_ 1^F$ \\ \hline
T2-2-3-E & $-1$ & $\psi \sim \mathbf{3}_ 0^F$ & $\phi (\varphi )\sim \mathbf{2}_ 1^S$  & $\Psi \sim \mathbf{2}_ 1^F$ \\ \hline
\multirow{2}{*}{T2-2-4-A} & $-1$ & --- & $\phi (\varphi )\sim \mathbf{2}_ {-1}^S$  & $\psi \sim \mathbf{1}_ {-2}^F,~\Psi \sim \mathbf{2}_ {-1}^F$ \\ \cline{2-5}
 & $1$ & $\psi \sim \mathbf{1}_ 0^F$ & $\phi (\varphi )\sim \mathbf{2}_ 1^S$ & $\Psi \sim \mathbf{2}_ {-1}^F$ \\ \hline
T2-2-4-B & $0$ & --- & $\phi (\varphi )\sim \mathbf{1}_ 0^S$ & $\psi \sim \mathbf{2}_ {-1}^F,~\Psi \sim \mathbf{2}_ {-1}^F$ \\ \hline
T2-2-4-E & $0$ & --- & $\phi (\varphi )\sim \mathbf{3}_ 0^S$  & $\psi \sim \mathbf{2}_ {-1}^F,~\Psi \sim \mathbf{2}_ {-1}^F$ \\ \hline
\multirow{2}{*}{T2-2-4-F} & $-1$ & --- & $\phi (\varphi )\sim \mathbf{2}_ {-1}^S$  & $\psi \sim \mathbf{3}_ {-2}^F,~\Psi \sim \mathbf{2}_ {-1}^F$ \\ \cline{2-5}
& $1$ & $\psi \sim \mathbf{3}_ 0^F$ & $\phi (\varphi )\sim \mathbf{2}_ 1^S$ & $\Psi \sim \mathbf{2}_ {-1}^F$ \\ \hline \hline
\end{tabular}
\caption{\label{tab:classify_model_min}The models that contain three different mediator fields with dark matter candidates.}
\end{table}

\section{Summary and conclusions}
\label{sec:conclusion}

Massive neutrinos is a well established experimental evidence for physics beyond SM, and the origin of tiny neutrino masses is a great puzzle of particle physics. The signal of neutrinoless double beta decay has not been observed so far, consequently neutrinos can be either Majorana or Dirac particles. Under the assumptions of Majorana neutrinos, the smallness of neutrino masses can be nicely explained through seesaw mechanism or can be radiatively generated. Both approaches have been extensively studied in the literature. For the case of Dirac neutrinos, the neutrino masses can be generated via the Dirac seesaw mechanism or generated through loop diagram.

One-loop realizations for the dimension four Dirac mass operator has been recently studied in~\cite{Ma:2016mwh,Wang:2016lve}. In the present work, we have systematically investigated the dimension six Dirac neutrino mass models at both tree level and one-loop level. We have identified all possible tree level and one-loop topologies. We find 13 one-loop diagrams in which both the tree diagrams in table~\ref{tab:tree_diagram} and Ma model can be avoided for certain quantum numbers of the mediators.
We have listed the new messenger fields and their transformation rules under the SM gauge symmetry, which are assumed to be scalars or fermions transforming as singlets, doublets or triplets of $SU(2)$. Moreover, we have presented the prediction for the neutrino mass matrix for each possible model. All these results are collected in tables~\ref{tab:figT2}\,--\,\ref{tab:figT8}. In order to ensure that the neutrino masses dominantly arise from the one-loop diagrams for the effective dimension six operator $\overline{\ell_{L}}\widetilde{H}\nu_{R}\left(H^{\dagger}H\right)$, one has to forbid the lower order contributions, in particular the direct Yukawa coupling    $\overline{\ell_{L}}\widetilde{H}\nu_{R}$ should disappear. We show that soft breaking abelian symmetry and finite non-abelian group can help to avoid the lower order neutrino masses. Note that the finite non-abelian group can also be used to explain the observed pattern of lepton mixing.

Another attractive feature of the radiative neutrino mass model is that
the internal messengers in the loop could be dark matter candidates. Taking into account the current bounds from direct detection experiments, the dark matter particle can be either scalar or fermion singlet and triplet with zero hypercharge, or a scalar doublet with $Y=\pm1$. For each viable model we have discussed the possible dark matter candidate and accordingly the parameter $\alpha$ is fixed to few discrete values. The results in tables~\ref{tab:figT2}\,--\,\ref{tab:figT8} provide new interesting opportunity for building models which can accommodate neutrino masses and dark matter simultaneously. In this work, we have presented the particle content and the dark matter candidates of each model. It is interesting to further perform detailed phenomenology study for some of these models, this will be left for future work.

\section*{Acknowledgements}
This work is supported by the National Natural Science Foundation of China
under Grant No.11522546.

\begin{appendix}

\clearpage

\begin{figure}[!htb]
\centering
\includegraphics[width=\linewidth]{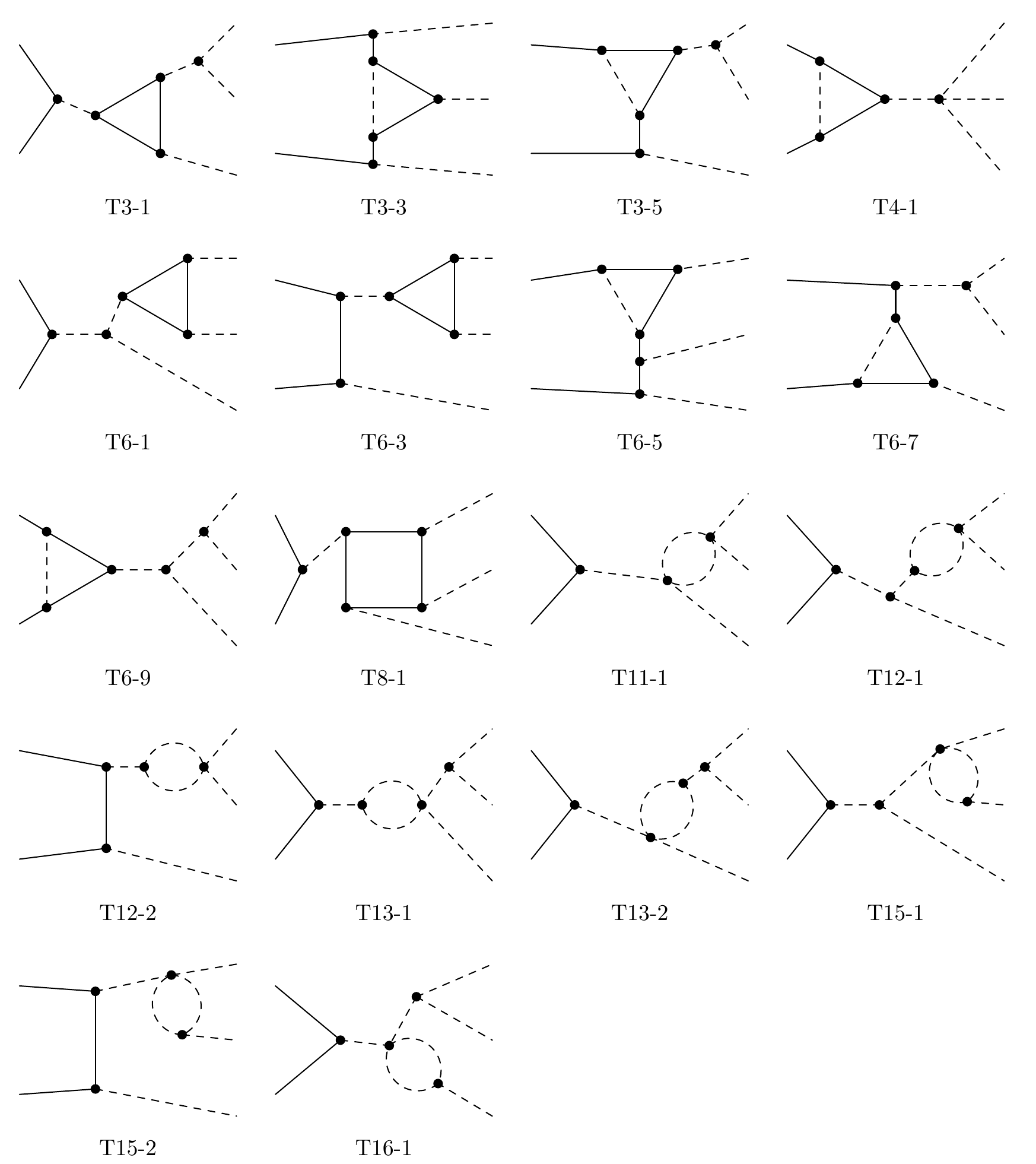}
\caption{\label{fig:divergent}Divergent one-loop Feynman diagrams for the dimension six operator $\overline{\ell_{L}}\widetilde{H}\nu_{R}\left(H^{\dagger}H\right)$. The dashed lines denote always scalars, and the solid lines denote fermions.}
\end{figure}

\begin{figure}[!htb]
  \centering
  \includegraphics[width=\linewidth]{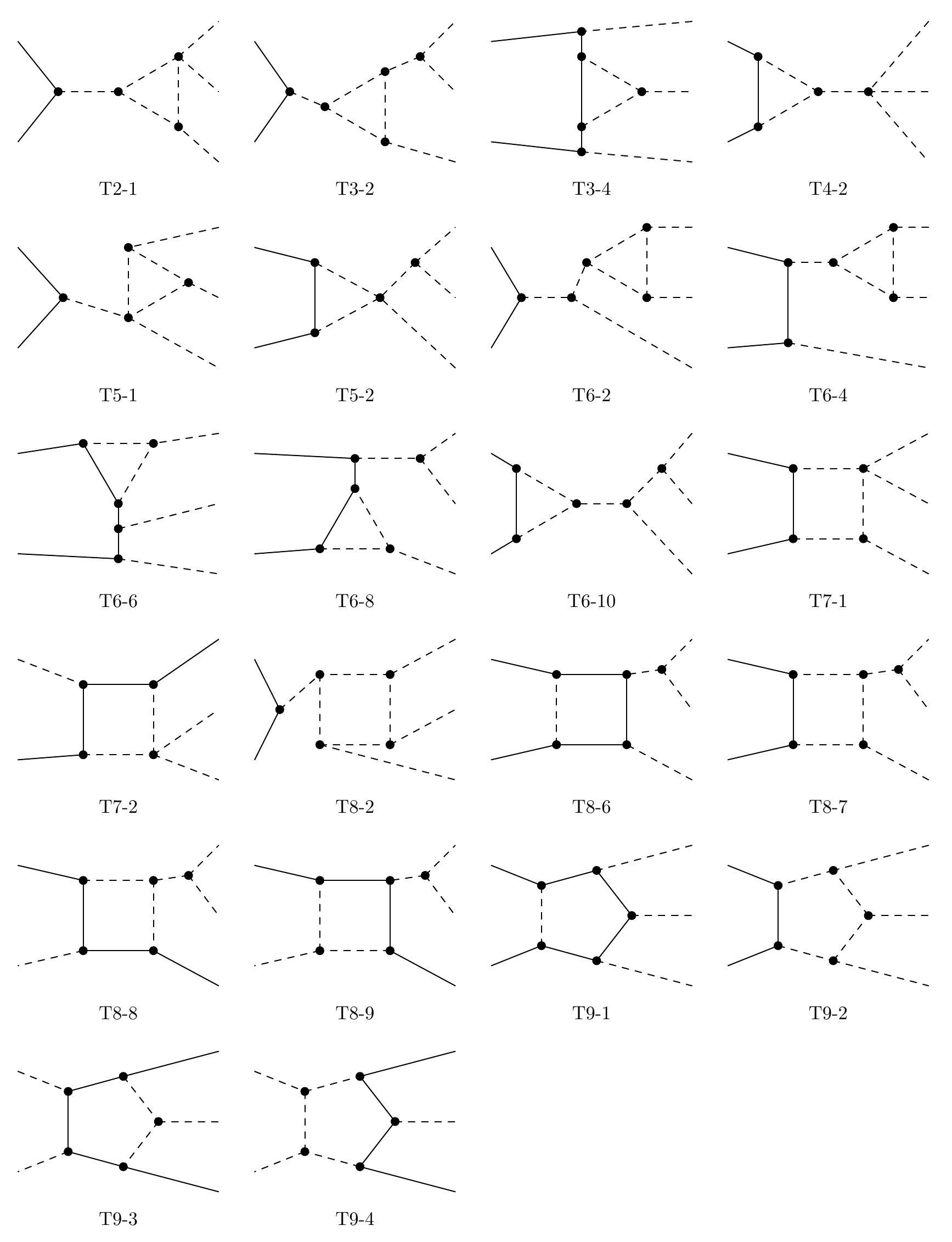}
\caption{\label{fig:top_FS_insertion_reducible}Finite one-loop diagrams of the neutrino mass operator $\overline{\ell_{L}}\widetilde{H}\nu_{R}\left(H^{\dagger}H\right)$, for which either the tree level realizations in table~\ref{tab:tree_diagram} or the one-loop diagrams in table~\ref{tab:Ma_model} are always present if no additional symmetry is introduced. The dashed lines denote scalars, and the solid lines denote fermions.
}
\end{figure}

\clearpage

\section{\label{sec:app_diagram}Field assignments, neutrino masses and dark matter candidates}

In this appendix, we show the 16 finite and irreducible diagrams which are not accompanied by the lower order tree level contributions in table~\ref{tab:tree_diagram} and Ma diagrams for certain quantum numbers of the mediators. These diagrams are generated from the topologies T2,
T3 and T8 shown in figure~\ref{fig:topfig}. We present the possible quantum number assignments for the mediators and the predictions for the neutrino masses, where the mediator fields are assumed to transform as singlets, doublets or triplets under $SU(2)$. If we require the neutrino masses dominantly arise from the following one-loop diagrams, both the tree level diagrams in table~\ref{tab:tree_diagram} and Ma diagrams in table~\ref{tab:Ma_model} should be avoided such that certain values of $\alpha$ are excluded. Moreover, we give the possible dark matter candidates and the corresponding values of $\alpha$ which are shown as subscript outside the square bracket. The dark matter particles are neutral components of the messengers in the loop. The $Z_2$ symmetry stabilizing dark matter can also help to forbid the tree level and Ma diagrams. As a consequence, many values of $\alpha$ excluded by lower order tree and Ma diagrams are allowed after the $Z_2$ symmetry is taken into account, and all these numbers are shadowed in grey in the following tables. Therefore we conclude that these one-loop models can simultaneously explain neutrino masses and accommodate dark matter.

\begin{longtable}{|m{4.5cm}<{\centering}|c|c|c|c|c|m{1.6cm}<{\centering}|m{1.6cm}<{\centering}|m{3.5cm}<{\centering}|N}\hline\hline
\endfirsthead

\hline
\endhead

\caption{(continued)}\\
\endfoot

\endlastfoot
\multirow{6}{*}{\includegraphics[width=\linewidth]{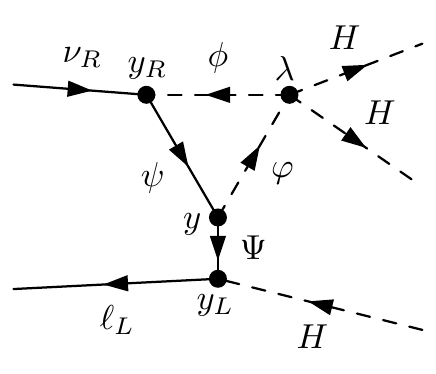}} & \multirow{2}{*}{} & \multirow{2}{*}{$\psi$} & \multirow{2}{*}{$\Psi$} & \multirow{2}{*}{$\phi$} & \multirow{2}{*}{$\varphi$} & \multicolumn{2}{c|}{\text{Excluded} $\alpha$} & \multirow{2}{*}{\text{Dark Matter}}\\ \cline{7-8}
 &  &  &  &  &  & \text{Tree} & \text{Ma} & \\ \cline{2-9}
 & A & $\mathbf{1}_{\alpha }^F$ & $\mathbf{3}_{-2}^F$ & $\mathbf{1}_{\alpha }^S$ & $\mathbf{3}_{\alpha +2}^S$ & \{\hl{$-4$}, \hl{$0$}\} & $\varnothing$ & $[\varphi ]_{-2}$, $[\psi ,\phi ,\varphi ]_{0}$ & \\[0.3cm] \cline{2-9}
 & B & $\mathbf{2}_{\alpha }^F$ & $\mathbf{3}_{-2}^F$ & $\mathbf{2}_{\alpha }^S$ & $\mathbf{2}_{\alpha +2}^S$ & \{\hl{$-3$}, \hl{$\pm 1$}\} & \{\hl{$\pm 3$}, \hl{$\pm 1$}\} & $[\varphi ]_{-3}$, $[\phi ,\varphi ]_{-1}^{\diamond}$, $[\phi ]_{1}$ & \\[0.3cm] \cline{2-9}
 & C & $\mathbf{3}_{\alpha }^F$ & $\mathbf{3}_{-2}^F$ & $\mathbf{3}_{\alpha }^S$ & $\mathbf{1}_{\alpha +2}^S$ & \{\hl{$\pm 2$}, \hl{$0$}\} & $\varnothing$ & $[\phi ,\varphi ]_{-2}$, $[\psi ,\phi ]_{0}$ & \\[0.3cm] \cline{2-9}
 T2-2-1 & D & $\mathbf{3}_{\alpha }^F$ & $\mathbf{3}_{-2}^F$ & $\mathbf{3}_{\alpha }^S$ & $\mathbf{3}_{\alpha +2}^S$ & \{\hl{$-4$}, \hl{$\pm 2$}, \hl{$0$}\} & $\varnothing$ & $[\phi ,\varphi ]_{-2}$, $[\psi ,\phi ,\varphi ]_{0}$\\ \hline
\multicolumn{9}{|c|}{$(m_{\nu})_{\alpha\beta}/\langle H\rangle^3=\frac{M_{\psi _j}}{M_{\Psi _i}}\lambda  {y_L}_{\alpha  i} y_{i j} {y_R}_{j \beta }I_{3}\left(M_{\phi },M_{\varphi },M_{\psi _j}\right)$}\\ \hline\hline
\multirow{7}{*}{\includegraphics[width=\linewidth]{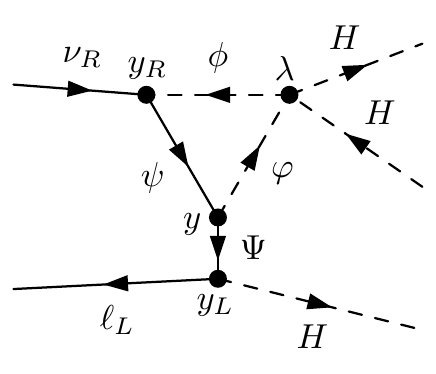}} & \multirow{2}{*}{} & \multirow{2}{*}{$\psi$} & \multirow{2}{*}{$\Psi$} & \multirow{2}{*}{$\phi$} & \multirow{2}{*}{$\varphi$} & \multicolumn{2}{c|}{\text{Excluded} $\alpha$} & \multirow{2}{*}{\text{Dark Matter}}\\ \cline{7-8}
 &  &  &  &  &  & \text{Tree} & \text{Ma} & \\ \cline{2-9}
 & A & $\mathbf{1}_{\alpha }^F$ & $\mathbf{1}_0^F$ & $\mathbf{1}_{\alpha }^S$ & $\mathbf{1}_{\alpha }^S$ & \{\hl{$0$}\} & $\varnothing$ & $[\psi ,\phi ,\varphi ]_{0}^{\diamond}$ & \\[0.1cm] \cline{2-9}
 & B & $\mathbf{1}_{\alpha }^F$ & $\mathbf{3}_0^F$ & $\mathbf{1}_{\alpha }^S$ & $\mathbf{3}_{\alpha }^S$ & \{\hl{$0$}\} & $\varnothing$ & $[\psi ,\phi ,\varphi ]_{0}$ & \\[0.1cm] \cline{2-9}
 & C & $\mathbf{2}_{\alpha }^F$ & $\mathbf{1}_0^F$ & $\mathbf{2}_{\alpha }^S$ & $\mathbf{2}_{\alpha }^S$ & \{\hl{$\pm 1$}\} & \{\hl{$\pm 1$}\} & $[\phi ,\varphi ]_{-1}^{\diamond}$, $[\phi ,\varphi ]_{1}^{\diamond}$ & \\[0.1cm] \cline{2-9}
 & D & $\mathbf{2}_{\alpha }^F$ & $\mathbf{3}_0^F$ & $\mathbf{2}_{\alpha }^S$ & $\mathbf{2}_{\alpha }^S$ & \{\hl{$\pm 1$}\} & \{\hl{$\pm 1$}\} & $[\phi ,\varphi ]_{-1}^{\diamond}$, $[\phi ,\varphi ]_{1}^{\diamond}$ & \\[0.1cm] \cline{2-9}
 & E & $\mathbf{3}_{\alpha }^F$ & $\mathbf{1}_0^F$ & $\mathbf{3}_{\alpha }^S$ & $\mathbf{3}_{\alpha }^S$ & \{\hl{$\pm 2$}, \hl{$0$}\} & $\varnothing$ & $[\psi ,\phi ,\varphi ]_{0}^{\diamond}$ & \\[0.1cm] \cline{2-9}
 & F & $\mathbf{3}_{\alpha }^F$ & $\mathbf{3}_0^F$ & $\mathbf{3}_{\alpha }^S$ & $\mathbf{1}_{\alpha }^S$ & \{\hl{$\pm 2$}, \hl{$0$}\} & $\varnothing$ & $[\psi ,\phi ,\varphi ]_{0}$ & \\[0.1cm] \cline{2-9}
 T2-2-2 & G & $\mathbf{3}_{\alpha }^F$ & $\mathbf{3}_0^F$ & $\mathbf{3}_{\alpha }^S$ & $\mathbf{3}_{\alpha }^S$ & \{\hl{$\pm 2$}, \hl{$0$}\} & $\varnothing$ & $[\psi ,\phi ,\varphi ]_{0}^{\diamond}$\\ \hline
\multicolumn{9}{|c|}{$(m_{\nu})_{\alpha\beta}/\langle H\rangle^3=\frac{M_{\psi _j}}{M_{\Psi _i}}\lambda  {y_L}_{\alpha  i} y_{i j} {y_R}_{j \beta }I_{3}\left(M_{\phi },M_{\varphi },M_{\psi _j}\right)$}\\ \hline\hline
\multirow{5}{*}{\includegraphics[width=\linewidth]{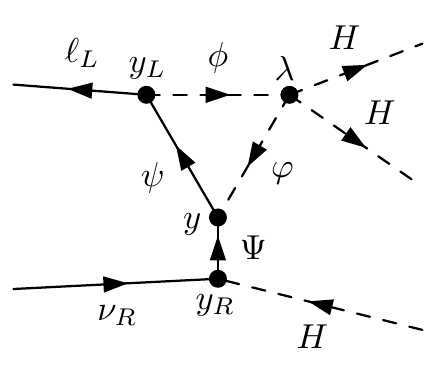}} & \multirow{2}{*}{} & \multirow{2}{*}{$\psi$} & \multirow{2}{*}{$\Psi$} & \multirow{2}{*}{$\phi$} & \multirow{2}{*}{$\varphi$} & \multicolumn{2}{c|}{\text{Excluded} $\alpha$} & \multirow{2}{*}{\text{Dark Matter}}\\ \cline{7-8}
 &  &  &  &  &  & \text{Tree} & \text{Ma} & \\ \cline{2-9}
 & A & $\mathbf{1}_{\alpha +1}^F$ & $\mathbf{2}_1^F$ & $\mathbf{2}_{\alpha +2}^S$ & $\mathbf{2}_{\alpha }^S$ & \{\hl{$-3$}, \hl{$\pm 1$}\} & \{\hl{$-3$}, \hl{$\pm 1$}\} & $[\phi ]_{-3}$, $[\psi ,\phi ,\varphi ]_{-1}^{\diamond}$, $[\varphi ]_{1}$\\ \cline{2-9}
 & B & $\mathbf{2}_{\alpha +1}^F$ & $\mathbf{2}_1^F$ & $\mathbf{1}_{\alpha +2}^S$ & $\mathbf{3}_{\alpha }^S$ & \{\hl{$\pm 2$}, \hl{$0$}\} & $\varnothing$ & $[\phi ,\varphi ]_{-2}$, $[\varphi ]_{0}$\\ \cline{2-9}
 & C & $\mathbf{2}_{\alpha +1}^F$ & $\mathbf{2}_1^F$ & $\mathbf{3}_{\alpha +2}^S$ & $\mathbf{1}_{\alpha }^S$ & \{\hl{$-4$}, \hl{$-2$}, \hl{$0$}\} & $\varnothing$ & $[\phi ]_{-2}$, $[\phi ,\varphi ]_{0}$\\ \cline{2-9}
 & D & $\mathbf{2}_{\alpha +1}^F$ & $\mathbf{2}_1^F$ & $\mathbf{3}_{\alpha +2}^S$ & $\mathbf{3}_{\alpha }^S$ & \{\hl{$-4$}, \hl{$\pm 2$}, \hl{$0$}\} & $\varnothing$ & $[\phi ,\varphi ]_{-2}$, $[\phi ,\varphi ]_{0}$\\ \cline{2-9}
 T2-2-3 & E & $\mathbf{3}_{\alpha +1}^F$ & $\mathbf{2}_1^F$ & $\mathbf{2}_{\alpha +2}^S$ & $\mathbf{2}_{\alpha }^S$ & \{\hl{$-3$}, \hl{$\pm 1$}\} & \{\hl{$-3$}, \hl{$\pm 1$}\} & $[\phi ]_{-3}$, $[\psi ,\phi ,\varphi ]_{-1}^{\diamond}$, $[\varphi ]_{1}$\\ \hline
\multicolumn{9}{|c|}{$(m_{\nu})_{\alpha\beta}/\langle H\rangle^3=\frac{M_{\psi _i}}{M_{\Psi _j}}\lambda  {y_L}_{\alpha  i} y_{i j} {y_R}_{j \beta }I_{3}\left(M_{\phi },M_{\varphi },M_{\psi _i}\right)$}\\ \hline\hline

%\newpage

\multirow{6}{*}{\includegraphics[width=\linewidth]{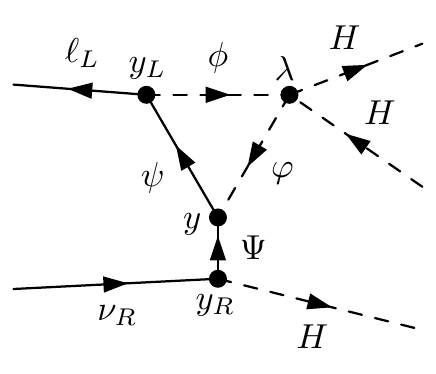}} & \multirow{2}{*}{} & \multirow{2}{*}{$\psi$} & \multirow{2}{*}{$\Psi$} & \multirow{2}{*}{$\phi$} & \multirow{2}{*}{$\varphi$} & \multicolumn{2}{c|}{\text{Excluded} $\alpha$} & \multirow{2}{*}{\text{Dark Matter}}\\ \cline{7-8}
 &  &  &  &  &  & \text{Tree} & \text{Ma} & \\ \cline{2-9}
 & A & $\mathbf{1}_{\alpha -1}^F$ & $\mathbf{2}_{-1}^F$ & $\mathbf{2}_{\alpha }^S$ & $\mathbf{2}_{\alpha }^S$ & \{\hl{$\pm 1$}\} & \{\hl{$\pm 1$}\} & $[\phi ,\varphi ]_{-1}^{\diamond}$, $[\psi ,\phi ,\varphi ]_{1}^{\diamond}$ & \\[0.1cm] \cline{2-9}
 & B & $\mathbf{2}_{\alpha -1}^F$ & $\mathbf{2}_{-1}^F$ & $\mathbf{1}_{\alpha }^S$ & $\mathbf{1}_{\alpha }^S$ & \{\hl{$0$}\} & $\varnothing$ & $[\phi ,\varphi ]_{0}^{\diamond}$ & \\[0.1cm] \cline{2-9}
 & C & $\mathbf{2}_{\alpha -1}^F$ & $\mathbf{2}_{-1}^F$ & $\mathbf{1}_{\alpha }^S$ & $\mathbf{3}_{\alpha }^S$ & \{\hl{$\pm 2$}, \hl{$0$}\} & $\varnothing$ & $[\phi ,\varphi ]_{0}$ & \\[0.1cm] \cline{2-9}
 & D & $\mathbf{2}_{\alpha -1}^F$ & $\mathbf{2}_{-1}^F$ & $\mathbf{3}_{\alpha }^S$ & $\mathbf{1}_{\alpha }^S$ & \{\hl{$\pm 2$}, \hl{$0$}\} & $\varnothing$ & $[\phi ,\varphi ]_{0}$ & \\[0.1cm] \cline{2-9}
 & E & $\mathbf{2}_{\alpha -1}^F$ & $\mathbf{2}_{-1}^F$ & $\mathbf{3}_{\alpha }^S$ & $\mathbf{3}_{\alpha }^S$ & \{\hl{$\pm 2$}, \hl{$0$}\} & $\varnothing$ & $[\phi ,\varphi ]_{0}^{\diamond}$ & \\[0.1cm] \cline{2-9}
T2-2-4 & F & $\mathbf{3}_{\alpha -1}^F$ & $\mathbf{2}_{-1}^F$ & $\mathbf{2}_{\alpha }^S$ & $\mathbf{2}_{\alpha }^S$ & \{\hl{$\pm 1$}, \hl{$3$}\} & \{\hl{$\pm 1$}\} & $[\phi ,\varphi ]_{-1}^{\diamond}$, $[\psi ,\phi ,\varphi ]_{1}^{\diamond}$ & \\[0.1cm] \hline
\multicolumn{9}{|c|}{$(m_{\nu})_{\alpha\beta}/\langle H\rangle^3=\frac{M_{\psi _i}}{M_{\Psi _j}}\lambda  {y_L}_{\alpha  i} y_{i j} {y_R}_{j \beta }I_{3}\left(M_{\phi },M_{\varphi },M_{\psi _i}\right)$}\\ \hline\hline
\caption{\label{tab:figT2}
The finite diagrams generated from the topology T2. We list the possible quantum number assignments for the mediators, the expressions for the neutrino mass matrix, and the dark matter candidates. We also display the values of $\alpha$ excluded from the appearance of tree level and Ma diagrams, where $\varnothing$ and $\mathbb{U}$ denote empty set and universal set respectively.}
 \end{longtable}

\begin{longtable}{|m{4.5cm}<{\centering}|c|c|c|c|c|c|m{1.6cm}<{\centering}|m{1.6cm}<{\centering}|m{3.5cm}<{\centering}|N}\hline\hline
\endfirsthead

\hline%\cline{2-9}
\endhead

\caption{(continued)}\\
\endfoot

\endlastfoot
\multirow{11}{*}{\includegraphics[width=\linewidth]{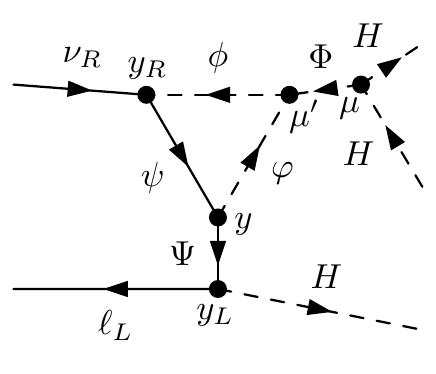}} & \multirow{2}{*}{} & \multirow{2}{*}{$\psi$} & \multirow{2}{*}{$\Psi$} & \multirow{2}{*}{$\phi$} & \multirow{2}{*}{$\varphi$} & \multirow{2}{*}{$\Phi$} & \multicolumn{2}{c|}{\text{Excluded} $\alpha$} & \multirow{2}{*}{\text{Dark Matter}}\\ \cline{8-9}
 &  &  &  &  &  &  & \text{Tree} & \text{Ma} & \\ \cline{2-10}
 & A & $\mathbf{1}_{\alpha }^F$ & $\mathbf{1}_0^F$ & $\mathbf{1}_{\alpha }^S$ & $\mathbf{1}_{\alpha }^S$ & $\mathbf{1}_0^S$ & $\mathbb{U}$ & $\varnothing$ & $[\psi ,\phi ,\varphi ]_{0}^{\diamond}$\\ \cline{2-10}
 & B & $\mathbf{1}_{\alpha }^F$ & $\mathbf{3}_0^F$ & $\mathbf{1}_{\alpha }^S$ & $\mathbf{3}_{\alpha }^S$ & $\mathbf{3}_0^S$ & $\mathbb{U}$ & $\varnothing$ & $[\psi ,\phi ,\varphi ]_{0}$\\ \cline{2-10}
 & C & $\mathbf{2}_{\alpha }^F$ & $\mathbf{1}_0^F$ & $\mathbf{2}_{\alpha }^S$ & $\mathbf{2}_{\alpha }^S$ & $\mathbf{1}_0^S$ & $\mathbb{U}$ & \{\hl{$\pm 1$}\} & $[\phi ,\varphi ]_{-1}^{\diamond}$, $[\phi ,\varphi ]_{1}^{\diamond}$\\ \cline{2-10}
 & D & $\mathbf{2}_{\alpha }^F$ & $\mathbf{1}_0^F$ & $\mathbf{2}_{\alpha }^S$ & $\mathbf{2}_{\alpha }^S$ & $\mathbf{3}_0^S$ & \{\hl{$\pm 1$}\} & \{\hl{$\pm 1$}\} & $[\phi ,\varphi ]_{-1}^{\diamond}$, $[\phi ,\varphi ]_{1}^{\diamond}$\\ \cline{2-10}
 & E & $\mathbf{2}_{\alpha }^F$ & $\mathbf{3}_0^F$ & $\mathbf{2}_{\alpha }^S$ & $\mathbf{2}_{\alpha }^S$ & $\mathbf{1}_0^S$ & \{\hl{$\pm 1$}\} & \{\hl{$\pm 1$}\} & $[\phi ,\varphi ]_{-1}^{\diamond}$, $[\phi ,\varphi ]_{1}^{\diamond}$\\ \cline{2-10}
 & F & $\mathbf{2}_{\alpha }^F$ & $\mathbf{3}_0^F$ & $\mathbf{2}_{\alpha }^S$ & $\mathbf{2}_{\alpha }^S$ & $\mathbf{3}_0^S$ & $\mathbb{U}$ & \{\hl{$\pm 1$}\} & $[\phi ,\varphi ]_{-1}^{\diamond}$, $[\phi ,\varphi ]_{1}^{\diamond}$\\ \cline{2-10}
 & G & $\mathbf{3}_{\alpha }^F$ & $\mathbf{1}_0^F$ & $\mathbf{3}_{\alpha }^S$ & $\mathbf{3}_{\alpha }^S$ & $\mathbf{1}_0^S$ & $\mathbb{U}$ & $\varnothing$ & $[\psi ,\phi ,\varphi ]_{0}^{\diamond}$\\ \cline{2-10}
 & H & $\mathbf{3}_{\alpha }^F$ & $\mathbf{1}_0^F$ & $\mathbf{3}_{\alpha }^S$ & $\mathbf{3}_{\alpha }^S$ & $\mathbf{3}_0^S$ & \{\hl{$\pm 2$}, \hl{$0$}\} & $\varnothing$ & $[\psi ,\phi ,\varphi ]_{0}^{\diamond}$\\ \cline{2-10}
 & I & $\mathbf{3}_{\alpha }^F$ & $\mathbf{3}_0^F$ & $\mathbf{3}_{\alpha }^S$ & $\mathbf{1}_{\alpha }^S$ & $\mathbf{3}_0^S$ & $\mathbb{U}$ & $\varnothing$ & $[\psi ,\phi ,\varphi ]_{0}$\\ \cline{2-10}
 & J & $\mathbf{3}_{\alpha }^F$ & $\mathbf{3}_0^F$ & $\mathbf{3}_{\alpha }^S$ & $\mathbf{3}_{\alpha }^S$ & $\mathbf{1}_0^S$ & \{\hl{$\pm 2$}, \hl{$0$}\} & $\varnothing$ & $[\psi ,\phi ,\varphi ]_{0}^{\diamond}$\\ \cline{2-10}
T3-6-1 & K & $\mathbf{3}_{\alpha }^F$ & $\mathbf{3}_0^F$ & $\mathbf{3}_{\alpha }^S$ & $\mathbf{3}_{\alpha }^S$ & $\mathbf{3}_0^S$ & $\mathbb{U}$ & $\varnothing$ & $[\psi ,\phi ,\varphi ]_{0}^{\diamond}$\\ \hline
 \multicolumn{10}{|c|}{$(m_{\nu})_{\alpha\beta}/\langle H\rangle^3=\frac{M_{\psi _j}}{M_{\Phi }^2 M_{\Psi _i}}\mu  \mu ' {y_L}_{\alpha  i} y_{i j} {y_R}_{j \beta }I_{3}\left(M_{\phi },M_{\varphi },M_{\psi _j}\right)$}\\ \hline\hline
\caption{\label{tab:figT3}
The finite diagrams generated from the topology T3. We list the possible quantum number assignments for the mediators, the expressions for the neutrino mass matrix, and the dark matter candidates. We also display the values of $\alpha$ excluded from the appearance of tree level and Ma diagrams, where $\varnothing$ and $\mathbb{U}$ denote empty set and universal set respectively.}
\end{longtable}

\newpage

\begin{longtable}{|m{4.5cm}<{\centering}|c|c|c|c|c|c|m{1.6cm}<{\centering}|m{1.6cm}<{\centering}|m{3.2cm}<{\centering}|}\hline\hline
\endfirsthead

\hline%\cline{2-9}
\endhead

\caption{(continued)}\\
\endfoot

\endlastfoot
\multirow{12}{*}{\includegraphics[width=\linewidth]{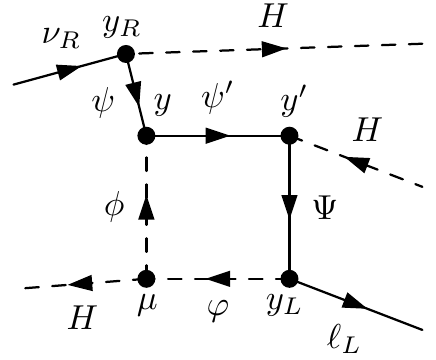}} & \multirow{2}{*}{} & \multirow{2}{*}{$\psi$} & \multirow{2}{*}{$\Psi$} & \multirow{2}{*}{$\psi '$} & \multirow{2}{*}{$\phi$} & \multirow{2}{*}{$\varphi$} & \multicolumn{2}{c|}{\text{Excluded} $\alpha$} & \multirow{2}{*}{\text{Dark Matter}}\\ \cline{8-9}
 &  &  &  &  &  &  & \text{Tree} & \text{Ma} & \\ \cline{2-10}
 & A & $\mathbf{2}_{-1}^F$ & $\mathbf{1}_{\alpha }^F$ & $\mathbf{2}_{\alpha -1}^F$ & $\mathbf{1}_{\alpha }^S$ & $\mathbf{2}_{\alpha +1}^S$ & \{\hl{$-2$}, \hl{$0$}\} & $\mathbb{U}$ & $[\varphi ]_{-2}$, $[\Psi ,\psi ',\phi ,\varphi ]_{0}$\\ \cline{2-10}
 & B & $\mathbf{2}_{-1}^F$ & $\mathbf{1}_{\alpha }^F$ & $\mathbf{2}_{\alpha -1}^F$ & $\mathbf{3}_{\alpha }^S$ & $\mathbf{2}_{\alpha +1}^S$ & \{\hl{$\pm 2$}, \hl{$0$}\} & \{\hl{$-2$}, $0$\} & $[\phi ,\varphi ]_{-2}$, $[\Psi ,\psi ',\phi ,\varphi ]_{0}$\\ \cline{2-10}
 & C & $\mathbf{2}_{-1}^F$ & $\mathbf{2}_{\alpha }^F$ & $\mathbf{1}_{\alpha -1}^F$ & $\mathbf{2}_{\alpha }^S$ & $\mathbf{1}_{\alpha +1}^S$ & \{\hl{$\pm 1$}\} & $\mathbb{U}$ & $[\phi ,\varphi ]_{-1}$, $[\Psi ,\psi ',\phi ]_{1}$\\ \cline{2-10}
 & D & $\mathbf{2}_{-1}^F$ & $\mathbf{2}_{\alpha }^F$ & $\mathbf{1}_{\alpha -1}^F$ & $\mathbf{2}_{\alpha }^S$ & $\mathbf{3}_{\alpha +1}^S$ & \{\hl{$-3$}, \hl{$\pm 1$}\} & $\mathbb{U}$ & $[\phi ,\varphi ]_{-1}$, $[\Psi ,\psi ',\phi ,\varphi ]_{1}$\\ \cline{2-10}
 & E & $\mathbf{2}_{-1}^F$ & $\mathbf{2}_{\alpha }^F$ & $\mathbf{3}_{\alpha -1}^F$ & $\mathbf{2}_{\alpha }^S$ & $\mathbf{1}_{\alpha +1}^S$ & \{\hl{$\pm 1$}, \hl{$3$}\} & $\mathbb{U}$ & $[\phi ,\varphi ]_{-1}$, $[\Psi ,\psi ',\phi ]_{1}$\\ \cline{2-10}
 & F & $\mathbf{2}_{-1}^F$ & $\mathbf{2}_{\alpha }^F$ & $\mathbf{3}_{\alpha -1}^F$ & $\mathbf{2}_{\alpha }^S$ & $\mathbf{3}_{\alpha +1}^S$ & \{\hl{$\pm 3$}, \hl{$\pm 1$}\} & $\mathbb{U}$ & $[\phi ,\varphi ]_{-1}$, $[\Psi ,\psi ',\phi ,\varphi ]_{1}$\\ \cline{2-10}
 T8-3-1 & G & $\mathbf{2}_{-1}^F$ & $\mathbf{3}_{\alpha }^F$ & $\mathbf{2}_{\alpha -1}^F$ & $\mathbf{1}_{\alpha }^S$ & $\mathbf{2}_{\alpha +1}^S$ & \{\hl{$\pm 2$}, \hl{$0$}\} & \{\hl{$-2$}, $0$\} & $[\varphi ]_{-2}$, $[\Psi ,\psi ',\phi ,\varphi ]_{0}$\\ \cline{2-10}
 & H & $\mathbf{2}_{-1}^F$ & $\mathbf{3}_{\alpha }^F$ & $\mathbf{2}_{\alpha -1}^F$ & $\mathbf{3}_{\alpha }^S$ & $\mathbf{2}_{\alpha +1}^S$ & \{\hl{$\pm 2$}, \hl{$0$}\} & $\mathbb{U}$ & $[\phi ,\varphi ]_{-2}$, $[\Psi ,\psi ',\phi ,\varphi ]_{0}$\\ \hline
\multicolumn{10}{|c|}{$(m_{\nu})_{\alpha\beta}/\langle H\rangle^3=-\frac{\mu }{M_{\psi _k}}{y_L}_{\alpha  i} y'_{i j} y_{j k} {y_R}_{k \beta }\left[M_{\Psi _i} M_{\psi '_j} I_{4}\left(M_{\phi },M_{\varphi },M_{\Psi _i},M_{\psi '_j}\right)+J_{4}\left(M_{\phi },M_{\varphi },M_{\Psi _i},M_{\psi '_j}\right)\right]$}\\ \hline\hline
\multirow{12}{*}{\includegraphics[width=\linewidth]{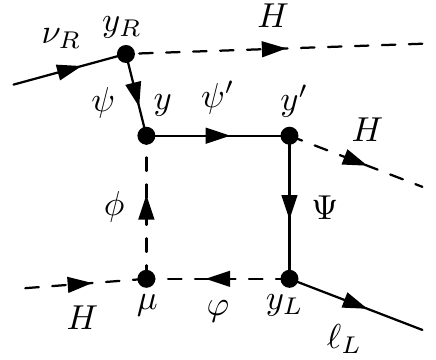}} & \multirow{2}{*}{} & \multirow{2}{*}{$\psi$} & \multirow{2}{*}{$\Psi$} & \multirow{2}{*}{$\psi '$} & \multirow{2}{*}{$\phi$} & \multirow{2}{*}{$\varphi$} & \multicolumn{2}{c|}{\text{Excluded} $\alpha$} & \multirow{2}{*}{\text{Dark Matter}}\\ \cline{8-9}
 &  &  &  &  &  &  & \text{Tree} & \text{Ma} & \\ \cline{2-10}
 & A & $\mathbf{2}_{-1}^F$ & $\mathbf{1}_{\alpha -1}^F$ & $\mathbf{2}_{\alpha }^F$ & $\mathbf{1}_{\alpha +1}^S$ & $\mathbf{2}_{\alpha }^S$ & \{\hl{$\pm 1$}\} & $\mathbb{U}$ & $[\phi ,\varphi ]_{-1}$, $[\Psi ,\psi ',\varphi ]_{1}$\\ \cline{2-10}
 & B & $\mathbf{2}_{-1}^F$ & $\mathbf{1}_{\alpha -1}^F$ & $\mathbf{2}_{\alpha }^F$ & $\mathbf{3}_{\alpha +1}^S$ & $\mathbf{2}_{\alpha }^S$ & \{\hl{$-3$}, \hl{$\pm 1$}\} & $\mathbb{U}$ & $[\phi ,\varphi ]_{-1}$, $[\Psi ,\psi ',\phi ,\varphi ]_{1}$\\ \cline{2-10}
 & C & $\mathbf{2}_{-1}^F$ & $\mathbf{2}_{\alpha -1}^F$ & $\mathbf{1}_{\alpha }^F$ & $\mathbf{2}_{\alpha +1}^S$ & $\mathbf{1}_{\alpha }^S$ & \{\hl{$-2$}, \hl{$0$}\} & $\mathbb{U}$ & $[\phi ]_{-2}$, $[\Psi ,\psi ',\phi ,\varphi ]_{0}$\\ \cline{2-10}
 & D & $\mathbf{2}_{-1}^F$ & $\mathbf{2}_{\alpha -1}^F$ & $\mathbf{1}_{\alpha }^F$ & $\mathbf{2}_{\alpha +1}^S$ & $\mathbf{3}_{\alpha }^S$ & \{\hl{$\pm 2$}, \hl{$0$}\} & \{\hl{$-2$}, $0$\} & $[\phi ,\varphi ]_{-2}$, $[\Psi ,\psi ',\phi ,\varphi ]_{0}$\\ \cline{2-10}
 & E & $\mathbf{2}_{-1}^F$ & $\mathbf{2}_{\alpha -1}^F$ & $\mathbf{3}_{\alpha }^F$ & $\mathbf{2}_{\alpha +1}^S$ & $\mathbf{1}_{\alpha }^S$ & \{\hl{$\pm 2$}, \hl{$0$}\} & \{\hl{$-2$}, $0$\} & $[\phi ]_{-2}$, $[\Psi ,\psi ',\phi ,\varphi ]_{0}$\\ \cline{2-10}
 & F & $\mathbf{2}_{-1}^F$ & $\mathbf{2}_{\alpha -1}^F$ & $\mathbf{3}_{\alpha }^F$ & $\mathbf{2}_{\alpha +1}^S$ & $\mathbf{3}_{\alpha }^S$ & \{\hl{$\pm 2$}, \hl{$0$}\} & $\mathbb{U}$ & $[\phi ,\varphi ]_{-2}$, $[\Psi ,\psi ',\phi ,\varphi ]_{0}$\\ \cline{2-10}
 T8-3-2 & G & $\mathbf{2}_{-1}^F$ & $\mathbf{3}_{\alpha -1}^F$ & $\mathbf{2}_{\alpha }^F$ & $\mathbf{1}_{\alpha +1}^S$ & $\mathbf{2}_{\alpha }^S$ & \{\hl{$\pm 1$}, \hl{$3$}\} & $\mathbb{U}$ & $[\phi ,\varphi ]_{-1}$, $[\Psi ,\psi ',\varphi ]_{1}$\\ \cline{2-10}
 & H & $\mathbf{2}_{-1}^F$ & $\mathbf{3}_{\alpha -1}^F$ & $\mathbf{2}_{\alpha }^F$ & $\mathbf{3}_{\alpha +1}^S$ & $\mathbf{2}_{\alpha }^S$ & \{\hl{$\pm 3$}, \hl{$\pm 1$}\} & $\mathbb{U}$ & $[\phi ,\varphi ]_{-1}$, $[\Psi ,\psi ',\phi ,\varphi ]_{1}$\\ \hline
\multicolumn{10}{|c|}{$(m_{\nu})_{\alpha\beta}/\langle H\rangle^3=-\frac{\mu }{M_{\psi _k}}{y_L}_{\alpha  i} y'_{i j} y_{j k} {y_R}_{k \beta }\left[M_{\Psi _i} M_{\psi '_j} I_{4}\left(M_{\phi },M_{\varphi },M_{\Psi _i},M_{\psi '_j}\right)+J_{4}\left(M_{\phi },M_{\varphi },M_{\Psi _i},M_{\psi '_j}\right)\right]$}\\ \hline\hline
\multirow{12}{*}{\includegraphics[width=\linewidth]{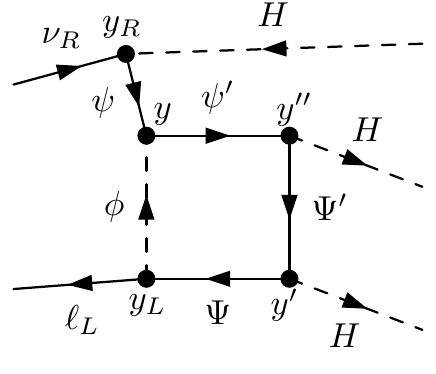}} & \multirow{2}{*}{} & \multirow{2}{*}{$\psi$} & \multirow{2}{*}{$\Psi$} & \multirow{2}{*}{$\psi '$} & \multirow{2}{*}{$\Psi '$} & \multirow{2}{*}{$\phi$} & \multicolumn{2}{c|}{\text{Excluded} $\alpha$} & \multirow{2}{*}{\text{Dark Matter}}\\ \cline{8-9}
 &  &  &  &  &  &  & \text{Tree} & \text{Ma} & \\ \cline{2-10}
 & A & $\mathbf{2}_1^F$ & $\mathbf{1}_{\alpha -1}^F$ & $\mathbf{1}_{\alpha +1}^F$ & $\mathbf{2}_{\alpha }^F$ & $\mathbf{2}_{\alpha }^S$ & \{\hl{$\pm 1$}\} & $\mathbb{U}$ & $[\psi ',\Psi ',\phi ]_{-1}$, $[\Psi ,\Psi ',\phi ]_{1}$\\ \cline{2-10}
 & B & $\mathbf{2}_1^F$ & $\mathbf{1}_{\alpha -1}^F$ & $\mathbf{3}_{\alpha +1}^F$ & $\mathbf{2}_{\alpha }^F$ & $\mathbf{2}_{\alpha }^S$ & \{\hl{$-3$}, \hl{$\pm 1$}\} & $\mathbb{U}$ & $[\psi ',\Psi ',\phi ]_{-1}$, $[\Psi ,\psi ',\Psi ',\phi ]_{1}$\\ \cline{2-10}
 & C & $\mathbf{2}_1^F$ & $\mathbf{2}_{\alpha -1}^F$ & $\mathbf{2}_{\alpha +1}^F$ & $\mathbf{1}_{\alpha }^F$ & $\mathbf{1}_{\alpha }^S$ & \{\hl{$0$}\} & $\mathbb{U}$ & $[\Psi ,\psi ',\Psi ',\phi ]_{0}^{\diamond}$\\ \cline{2-10}
 & D & $\mathbf{2}_1^F$ & $\mathbf{2}_{\alpha -1}^F$ & $\mathbf{2}_{\alpha +1}^F$ & $\mathbf{1}_{\alpha }^F$ & $\mathbf{3}_{\alpha }^S$ & \{\hl{$\pm 2$}, \hl{$0$}\} & $\varnothing$ & $[\Psi ,\psi ',\Psi ',\phi ]_{0}^{\diamond}$\\ \cline{2-10}
 & E & $\mathbf{2}_1^F$ & $\mathbf{2}_{\alpha -1}^F$ & $\mathbf{2}_{\alpha +1}^F$ & $\mathbf{3}_{\alpha }^F$ & $\mathbf{1}_{\alpha }^S$ & \{\hl{$\pm 2$}, \hl{$0$}\} & $\varnothing$ & $[\Psi ,\psi ',\Psi ',\phi ]_{0}^{\diamond}$\\ \cline{2-10}
 & F & $\mathbf{2}_1^F$ & $\mathbf{2}_{\alpha -1}^F$ & $\mathbf{2}_{\alpha +1}^F$ & $\mathbf{3}_{\alpha }^F$ & $\mathbf{3}_{\alpha }^S$ & \{\hl{$\pm 2$}, \hl{$0$}\} & $\mathbb{U}$ & $[\Psi ,\psi ',\Psi ',\phi ]_{0}^{\diamond}$\\ \cline{2-10}
 T8-4-1 & G & $\mathbf{2}_1^F$ & $\mathbf{3}_{\alpha -1}^F$ & $\mathbf{1}_{\alpha +1}^F$ & $\mathbf{2}_{\alpha }^F$ & $\mathbf{2}_{\alpha }^S$ & \{\hl{$\pm 1$}, \hl{$3$}\} & $\mathbb{U}$ & $[\Psi ,\psi ',\Psi ',\phi ]_{-1}$, $[\Psi ,\Psi ',\phi ]_{1}$\\ \cline{2-10}
 & H & $\mathbf{2}_1^F$ & $\mathbf{3}_{\alpha -1}^F$ & $\mathbf{3}_{\alpha +1}^F$ & $\mathbf{2}_{\alpha }^F$ & $\mathbf{2}_{\alpha }^S$ & \{\hl{$\pm 3$}, \hl{$\pm 1$}\} & $\mathbb{U}$ & $[\Psi ,\psi ',\Psi ',\phi ]_{-1}$, $[\Psi ,\psi ',\Psi ',\phi ]_{1}$\\ \hline
 \multicolumn{10}{|m{\linewidth}<{\centering}|}{$(m_{\nu})_{\alpha\beta}/\langle H\rangle^3=-\frac{{y_L}_{\alpha  i} y'_{i j} y''_{j k} y_{k l} {y_R}_{l \beta }}{M_{\psi _l}}\left[M_{\Psi _i} M_{\Psi '_j} M_{\psi '_k} I_{4}\left(M_{\phi },M_{\Psi _i},M_{\psi '_k},M_{\Psi '_j}\right)\right.$ $\left.+\left(M_{\Psi _i}+M_{\Psi '_j}+M_{\psi '_k}\right) J_{4}\left(M_{\phi },M_{\Psi _i},M_{\psi '_k},M_{\Psi '_j}\right)\right]$}\\ \hline\hline

\newpage

 \multirow{12}{*}{\includegraphics[width=\linewidth]{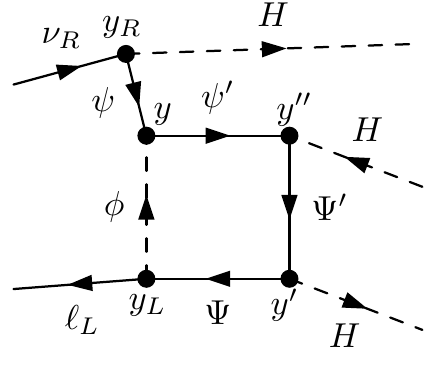}} & \multirow{2}{*}{} & \multirow{2}{*}{$\psi$} & \multirow{2}{*}{$\Psi$} & \multirow{2}{*}{$\psi '$} & \multirow{2}{*}{$\Psi '$} & \multirow{2}{*}{$\phi$} & \multicolumn{2}{c|}{\text{Excluded} $\alpha$} & \multirow{2}{*}{\text{Dark Matter}}\\ \cline{8-9}
 &  &  &  &  &  &  & \text{Tree} & \text{Ma} & \\ \cline{2-10}
 & A & $\mathbf{2}_{-1}^F$ & $\mathbf{1}_{\alpha }^F$ & $\mathbf{1}_{\alpha }^F$ & $\mathbf{2}_{\alpha +1}^F$ & $\mathbf{2}_{\alpha +1}^S$ & \{\hl{$-2$}, \hl{$0$}\} & $\mathbb{U}$ & $[\phi ]_{-2}^{\diamond}$, $[\Psi ,\psi ',\Psi ',\phi ]_{0}^{\diamond}$\\ \cline{2-10}
 & B & $\mathbf{2}_{-1}^F$ & $\mathbf{1}_{\alpha }^F$ & $\mathbf{3}_{\alpha }^F$ & $\mathbf{2}_{\alpha +1}^F$ & $\mathbf{2}_{\alpha +1}^S$ & \{\hl{$\pm 2$}, \hl{$0$}\} & $\mathbb{U}$ & $[\phi ]_{-2}$, $[\Psi ,\psi ',\Psi ',\phi ]_{0}$\\ \cline{2-10}
 & C & $\mathbf{2}_{-1}^F$ & $\mathbf{2}_{\alpha }^F$ & $\mathbf{2}_{\alpha }^F$ & $\mathbf{1}_{\alpha +1}^F$ & $\mathbf{1}_{\alpha +1}^S$ & \{\hl{$-1$}\} & $\mathbb{U}$ & $[\Psi ,\psi ',\Psi ',\phi ]_{-1}^{\diamond}$\\ \cline{2-10}
 & D & $\mathbf{2}_{-1}^F$ & $\mathbf{2}_{\alpha }^F$ & $\mathbf{2}_{\alpha }^F$ & $\mathbf{1}_{\alpha +1}^F$ & $\mathbf{3}_{\alpha +1}^S$ & \{\hl{$-3$}, \hl{$\pm 1$}\} & $\varnothing$ & $[\Psi ,\psi ',\Psi ',\phi ]_{-1}^{\diamond}$\\ \cline{2-10}
 & E & $\mathbf{2}_{-1}^F$ & $\mathbf{2}_{\alpha }^F$ & $\mathbf{2}_{\alpha }^F$ & $\mathbf{3}_{\alpha +1}^F$ & $\mathbf{1}_{\alpha +1}^S$ & \{\hl{$-3$}, \hl{$\pm 1$}\} & $\varnothing$ & $[\Psi ,\psi ',\Psi ',\phi ]_{-1}^{\diamond}$\\ \cline{2-10}
 & F & $\mathbf{2}_{-1}^F$ & $\mathbf{2}_{\alpha }^F$ & $\mathbf{2}_{\alpha }^F$ & $\mathbf{3}_{\alpha +1}^F$ & $\mathbf{3}_{\alpha +1}^S$ & \{\hl{$-3$}, \hl{$\pm 1$}\} & $\mathbb{U}$ & $[\Psi ,\psi ',\Psi ',\phi ]_{-1}^{\diamond}$\\ \cline{2-10}
 T8-4-2 & G & $\mathbf{2}_{-1}^F$ & $\mathbf{3}_{\alpha }^F$ & $\mathbf{1}_{\alpha }^F$ & $\mathbf{2}_{\alpha +1}^F$ & $\mathbf{2}_{\alpha +1}^S$ & \{\hl{$\pm 2$}, \hl{$0$}\} & $\mathbb{U}$ & $[\phi ]_{-2}$, $[\Psi ,\psi ',\Psi ',\phi ]_{0}$\\ \cline{2-10}
 & H & $\mathbf{2}_{-1}^F$ & $\mathbf{3}_{\alpha }^F$ & $\mathbf{3}_{\alpha }^F$ & $\mathbf{2}_{\alpha +1}^F$ & $\mathbf{2}_{\alpha +1}^S$ & \{\hl{$\pm 2$}, \hl{$0$}\} & $\mathbb{U}$ & $[\phi ]_{-2}^{\diamond}$, $[\Psi ,\psi ',\Psi ',\phi ]_{0}^{\diamond}$\\ \hline
\multicolumn{10}{|m{\linewidth}<{\centering}|}{$(m_{\nu})_{\alpha\beta}/\langle H\rangle^3=-\frac{{y_L}_{\alpha  i} y'_{i j} y''_{j k} y_{k l} {y_R}_{l \beta }}{M_{\psi _l}}\left[M_{\Psi _i} M_{\Psi '_j} M_{\psi '_k} I_{4}\left(M_{\phi },M_{\Psi _i},M_{\psi '_k},M_{\Psi '_j}\right)\right.$ $\left.+\left(M_{\Psi _i}+M_{\Psi '_j}+M_{\psi '_k}\right) J_{4}\left(M_{\phi },M_{\Psi _i},M_{\psi '_k},M_{\Psi '_j}\right)\right]$}\\ \hline\hline
\multirow{12}{*}{\includegraphics[width=\linewidth]{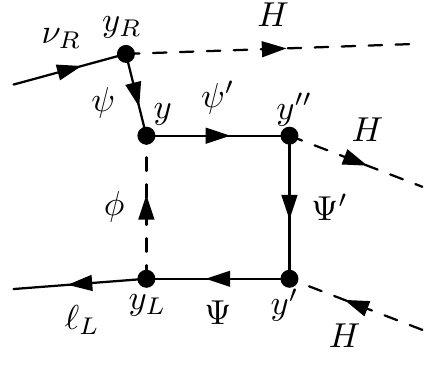}} & \multirow{2}{*}{} & \multirow{2}{*}{$\psi$} & \multirow{2}{*}{$\Psi$} & \multirow{2}{*}{$\psi '$} & \multirow{2}{*}{$\Psi '$} & \multirow{2}{*}{$\phi$} & \multicolumn{2}{c|}{\text{Excluded} $\alpha$} & \multirow{2}{*}{\text{Dark Matter}}\\ \cline{8-9}
 &  &  &  &  &  &  & \text{Tree} & \text{Ma} & \\ \cline{2-10}
 & A & $\mathbf{2}_{-1}^F$ & $\mathbf{1}_{\alpha }^F$ & $\mathbf{1}_{\alpha }^F$ & $\mathbf{2}_{\alpha -1}^F$ & $\mathbf{2}_{\alpha +1}^S$ & \{\hl{$-2$}, \hl{$0$}\} & \{\hl{$-2$}, $0$\} & $[\phi ]_{-2}^{\diamond}$, $[\Psi ,\psi ',\Psi ',\phi ]_{0}^{\diamond}$\\ \cline{2-10}
 & B & $\mathbf{2}_{-1}^F$ & $\mathbf{1}_{\alpha }^F$ & $\mathbf{3}_{\alpha }^F$ & $\mathbf{2}_{\alpha -1}^F$ & $\mathbf{2}_{\alpha +1}^S$ & \{\hl{$\pm 2$}, \hl{$0$}\} & \{\hl{$-2$}, $0$\} & $[\phi ]_{-2}$, $[\Psi ,\psi ',\Psi ',\phi ]_{0}$\\ \cline{2-10}
 & C & $\mathbf{2}_{-1}^F$ & $\mathbf{2}_{\alpha }^F$ & $\mathbf{2}_{\alpha }^F$ & $\mathbf{1}_{\alpha -1}^F$ & $\mathbf{1}_{\alpha +1}^S$ & \{\hl{$\pm 1$}\} & \{$0$\} & $[\phi ]_{-1}^{\diamond}$, $[\Psi ,\psi ',\Psi ']_{1}^{\diamond}$\\ \cline{2-10}
 & D & $\mathbf{2}_{-1}^F$ & $\mathbf{2}_{\alpha }^F$ & $\mathbf{2}_{\alpha }^F$ & $\mathbf{1}_{\alpha -1}^F$ & $\mathbf{3}_{\alpha +1}^S$ & \{\hl{$-3$}, \hl{$\pm 1$}\} & $\varnothing$ & $[\phi ]_{-1}^{\diamond}$, $[\Psi ,\psi ',\Psi ']_{1}^{\diamond}$\\ \cline{2-10}
 & E & $\mathbf{2}_{-1}^F$ & $\mathbf{2}_{\alpha }^F$ & $\mathbf{2}_{\alpha }^F$ & $\mathbf{3}_{\alpha -1}^F$ & $\mathbf{1}_{\alpha +1}^S$ & \{\hl{$\pm 1$}, \hl{$3$}\} & $\varnothing$ & $[\phi ]_{-1}^{\diamond}$, $[\Psi ,\psi ',\Psi ']_{1}^{\diamond}$\\ \cline{2-10}
 & F & $\mathbf{2}_{-1}^F$ & $\mathbf{2}_{\alpha }^F$ & $\mathbf{2}_{\alpha }^F$ & $\mathbf{3}_{\alpha -1}^F$ & $\mathbf{3}_{\alpha +1}^S$ & \{\hl{$\pm 3$}, \hl{$\pm 1$}\} & \{$0$\} & $[\phi ]_{-1}^{\diamond}$, $[\Psi ,\psi ',\Psi ']_{1}^{\diamond}$\\ \cline{2-10}
 T8-4-3 & G & $\mathbf{2}_{-1}^F$ & $\mathbf{3}_{\alpha }^F$ & $\mathbf{1}_{\alpha }^F$ & $\mathbf{2}_{\alpha -1}^F$ & $\mathbf{2}_{\alpha +1}^S$ & \{\hl{$\pm 2$}, \hl{$0$}\} & \{\hl{$-2$}, $0$\} & $[\phi ]_{-2}$, $[\Psi ,\psi ',\Psi ',\phi ]_{0}$\\ \cline{2-10}
 & H & $\mathbf{2}_{-1}^F$ & $\mathbf{3}_{\alpha }^F$ & $\mathbf{3}_{\alpha }^F$ & $\mathbf{2}_{\alpha -1}^F$ & $\mathbf{2}_{\alpha +1}^S$ & \{\hl{$\pm 2$}, \hl{$0$}\} & \{\hl{$-2$}, $0$\} & $[\phi ]_{-2}^{\diamond}$, $[\Psi ,\psi ',\Psi ',\phi ]_{0}^{\diamond}$\\ \hline
\multicolumn{10}{|m{\linewidth}<{\centering}|}{$(m_{\nu})_{\alpha\beta}/\langle H\rangle^3=-\frac{{y_L}_{\alpha  i} y'_{i j} y''_{j k} y_{k l} {y_R}_{l \beta }}{M_{\psi _l}}\left[M_{\Psi _i} M_{\Psi '_j} M_{\psi '_k} I_{4}\left(M_{\phi },M_{\Psi _i},M_{\psi '_k},M_{\Psi '_j}\right)\right.$ $\left.+\left(M_{\Psi _i}+M_{\Psi '_j}+M_{\psi '_k}\right) J_{4}\left(M_{\phi },M_{\Psi _i},M_{\psi '_k},M_{\Psi '_j}\right)\right]$}\\ \hline\hline
\multirow{14}{*}{\includegraphics[width=\linewidth]{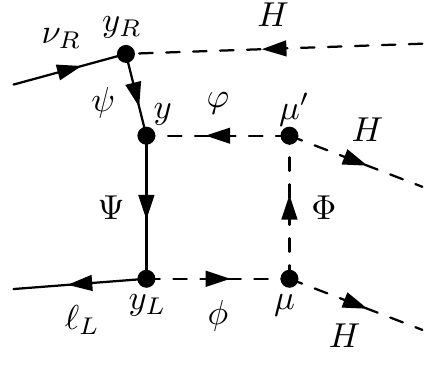}} & \multirow{2}{*}{} & \multirow{2}{*}{$\psi$} & \multirow{2}{*}{$\Psi$} & \multirow{2}{*}{$\phi$} & \multirow{2}{*}{$\varphi$} & \multirow{2}{*}{$\Phi$} & \multicolumn{2}{c|}{\text{Excluded} $\alpha$} & \multirow{2}{*}{\text{Dark Matter}}\\ \cline{8-9}
 &  &  &  &  &  &  & \text{Tree} & \text{Ma} & \\ \cline{2-10}
 & A & $\mathbf{2}_1^F$ & $\mathbf{1}_{\alpha }^F$ & $\mathbf{2}_{\alpha +1}^S$ & $\mathbf{2}_{\alpha -1}^S$ & $\mathbf{1}_{\alpha }^S$ & \{\hl{$\pm 2$}, \hl{$0$}\} & $\mathbb{U}$ & $[\phi ]_{-2}$, $[\Psi ,\phi ,\varphi ,\Phi ]_{0}^{\diamond}$, $[\varphi ]_{2}$\\ \cline{2-10}
 & B & $\mathbf{2}_1^F$ & $\mathbf{1}_{\alpha }^F$ & $\mathbf{2}_{\alpha +1}^S$ & $\mathbf{2}_{\alpha -1}^S$ & $\mathbf{3}_{\alpha }^S$ & \{\hl{$\pm 2$}, \hl{$0$}\} & \{\hl{$\pm 2$}, \hl{$0$}\} & $[\phi ,\Phi ]_{-2}$, $[\Psi ,\phi ,\varphi ,\Phi ]_{0}^{\diamond}$, $[\varphi ,\Phi ]_{2}$\\ \cline{2-10}
 & C & $\mathbf{2}_1^F$ & $\mathbf{2}_{\alpha }^F$ & $\mathbf{1}_{\alpha +1}^S$ & $\mathbf{1}_{\alpha -1}^S$ & $\mathbf{2}_{\alpha }^S$ & \{\hl{$\pm 1$}\} & $\mathbb{U}$ & $[\phi ,\Phi ]_{-1}$, $[\varphi ,\Phi ]_{1}$\\ \cline{2-10}
 & D & $\mathbf{2}_1^F$ & $\mathbf{2}_{\alpha }^F$ & $\mathbf{1}_{\alpha +1}^S$ & $\mathbf{3}_{\alpha -1}^S$ & $\mathbf{2}_{\alpha }^S$ & \{\hl{$\pm 1$}, \hl{$3$}\} & $\mathbb{U}$ & $[\phi ,\varphi ,\Phi ]_{-1}$, $[\varphi ,\Phi ]_{1}$\\ \cline{2-10}
 & E & $\mathbf{2}_1^F$ & $\mathbf{2}_{\alpha }^F$ & $\mathbf{3}_{\alpha +1}^S$ & $\mathbf{1}_{\alpha -1}^S$ & $\mathbf{2}_{\alpha }^S$ & \{\hl{$-3$}, \hl{$\pm 1$}\} & $\mathbb{U}$ & $[\phi ,\Phi ]_{-1}$, $[\phi ,\varphi ,\Phi ]_{1}$\\ \cline{2-10}
 & F & $\mathbf{2}_1^F$ & $\mathbf{2}_{\alpha }^F$ & $\mathbf{3}_{\alpha +1}^S$ & $\mathbf{3}_{\alpha -1}^S$ & $\mathbf{2}_{\alpha }^S$ & \{\hl{$\pm 3$}, \hl{$\pm 1$}\} & $\mathbb{U}$ & $[\phi ,\varphi ,\Phi ]_{-1}$, $[\phi ,\varphi ,\Phi ]_{1}$\\ \cline{2-10}
 T8-5-1 & G & $\mathbf{2}_1^F$ & $\mathbf{3}_{\alpha }^F$ & $\mathbf{2}_{\alpha +1}^S$ & $\mathbf{2}_{\alpha -1}^S$ & $\mathbf{1}_{\alpha }^S$ & \{\hl{$\pm 2$}, \hl{$0$}\} & \{\hl{$\pm 2$}, \hl{$0$}\} & $[\phi ]_{-2}$, $[\Psi ,\phi ,\varphi ,\Phi ]_{0}^{\diamond}$, $[\varphi ]_{2}$\\ \cline{2-10}
 & H & $\mathbf{2}_1^F$ & $\mathbf{3}_{\alpha }^F$ & $\mathbf{2}_{\alpha +1}^S$ & $\mathbf{2}_{\alpha -1}^S$ & $\mathbf{3}_{\alpha }^S$ & \{\hl{$\pm 2$}, \hl{$0$}\} & $\mathbb{U}$ & $[\phi ,\Phi ]_{-2}$, $[\Psi ,\phi ,\varphi ,\Phi ]_{0}^{\diamond}$, $[\varphi ,\Phi ]_{2}$\\ \hline
\multicolumn{10}{|c|}{$(m_{\nu})_{\alpha\beta}/\langle H\rangle^3=-\frac{M_{\Psi _i}}{M_{\psi _j}}\mu  \mu ' {y_L}_{\alpha  i} y_{i j} {y_R}_{j \beta }I_{4}\left(M_{\phi },M_{\varphi },M_{\Phi },M_{\Psi _i}\right)$}\\ \hline\hline
\multirow{14}{*}{\includegraphics[width=\linewidth]{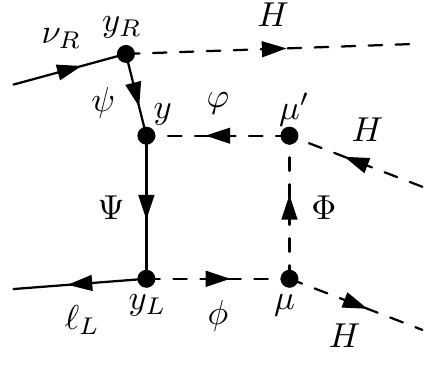}} & \multirow{2}{*}{} & \multirow{2}{*}{$\psi$} & \multirow{2}{*}{$\Psi$} & \multirow{2}{*}{$\phi$} & \multirow{2}{*}{$\varphi$} & \multirow{2}{*}{$\Phi$} & \multicolumn{2}{c|}{\text{Excluded} $\alpha$} & \multirow{2}{*}{\text{Dark Matter}}\\ \cline{8-9}
 &  &  &  &  &  &  & \text{Tree} & \text{Ma} & \\ \cline{2-10}
 & A & $\mathbf{2}_{-1}^F$ & $\mathbf{1}_{\alpha }^F$ & $\mathbf{2}_{\alpha +1}^S$ & $\mathbf{2}_{\alpha +1}^S$ & $\mathbf{1}_{\alpha }^S$ & \{\hl{$-2$}, \hl{$0$}\} & $\mathbb{U}$ & $[\phi ,\varphi ]_{-2}^{\diamond}$, $[\Psi ,\phi ,\varphi ,\Phi ]_{0}^{\diamond}$\\ \cline{2-10}
 & B & $\mathbf{2}_{-1}^F$ & $\mathbf{1}_{\alpha }^F$ & $\mathbf{2}_{\alpha +1}^S$ & $\mathbf{2}_{\alpha +1}^S$ & $\mathbf{3}_{\alpha }^S$ & \{\hl{$\pm 2$}, \hl{$0$}\} & \{\hl{$-2$}, \hl{$0$}\} & $[\phi ,\varphi ,\Phi ]_{-2}^{\diamond}$, $[\Psi ,\phi ,\varphi ,\Phi ]_{0}^{\diamond}$\\ \cline{2-10}
 & C & $\mathbf{2}_{-1}^F$ & $\mathbf{2}_{\alpha }^F$ & $\mathbf{1}_{\alpha +1}^S$ & $\mathbf{1}_{\alpha +1}^S$ & $\mathbf{2}_{\alpha }^S$ & \{\hl{$\pm 1$}\} & $\mathbb{U}$ & $[\phi ,\varphi ,\Phi ]_{-1}^{\diamond}$, $[\Phi ]_{1}^{\diamond}$\\ \cline{2-10}
 & D & $\mathbf{2}_{-1}^F$ & $\mathbf{2}_{\alpha }^F$ & $\mathbf{1}_{\alpha +1}^S$ & $\mathbf{3}_{\alpha +1}^S$ & $\mathbf{2}_{\alpha }^S$ & \{\hl{$-3$}, \hl{$\pm 1$}\} & $\mathbb{U}$ & $[\phi ,\varphi ,\Phi ]_{-1}$, $[\varphi ,\Phi ]_{1}$\\ \cline{2-10}
 & E & $\mathbf{2}_{-1}^F$ & $\mathbf{2}_{\alpha }^F$ & $\mathbf{3}_{\alpha +1}^S$ & $\mathbf{1}_{\alpha +1}^S$ & $\mathbf{2}_{\alpha }^S$ & \{\hl{$-3$}, \hl{$\pm 1$}\} & $\mathbb{U}$ & $[\phi ,\varphi ,\Phi ]_{-1}$, $[\phi ,\Phi ]_{1}$\\ \cline{2-10}
 & F & $\mathbf{2}_{-1}^F$ & $\mathbf{2}_{\alpha }^F$ & $\mathbf{3}_{\alpha +1}^S$ & $\mathbf{3}_{\alpha +1}^S$ & $\mathbf{2}_{\alpha }^S$ & \{\hl{$-3$}, \hl{$\pm 1$}\} & $\mathbb{U}$ & $[\phi ,\varphi ,\Phi ]_{-1}^{\diamond}$, $[\phi ,\varphi ,\Phi ]_{1}^{\diamond}$\\ \cline{2-10}
 T8-5-2 & G & $\mathbf{2}_{-1}^F$ & $\mathbf{3}_{\alpha }^F$ & $\mathbf{2}_{\alpha +1}^S$ & $\mathbf{2}_{\alpha +1}^S$ & $\mathbf{1}_{\alpha }^S$ & \{\hl{$\pm 2$}, \hl{$0$}\} & \{\hl{$-2$}, \hl{$0$}\} & $[\phi ,\varphi ]_{-2}^{\diamond}$, $[\Psi ,\phi ,\varphi ,\Phi ]_{0}^{\diamond}$\\ \cline{2-10}
 & H & $\mathbf{2}_{-1}^F$ & $\mathbf{3}_{\alpha }^F$ & $\mathbf{2}_{\alpha +1}^S$ & $\mathbf{2}_{\alpha +1}^S$ & $\mathbf{3}_{\alpha }^S$ & \{\hl{$\pm 2$}, \hl{$0$}\} & $\mathbb{U}$ & $[\phi ,\varphi ,\Phi ]_{-2}^{\diamond}$, $[\Psi ,\phi ,\varphi ,\Phi ]_{0}^{\diamond}$\\ \hline
\multicolumn{10}{|c|}{$(m_{\nu})_{\alpha\beta}/\langle H\rangle^3=-\frac{M_{\Psi _i}}{M_{\psi _j}}\mu  \mu ' {y_L}_{\alpha  i} y_{i j} {y_R}_{j \beta }I_{4}\left(M_{\phi },M_{\varphi },M_{\Phi },M_{\Psi _i}\right)$}\\ \hline\hline
\multirow{14}{*}{\includegraphics[width=\linewidth]{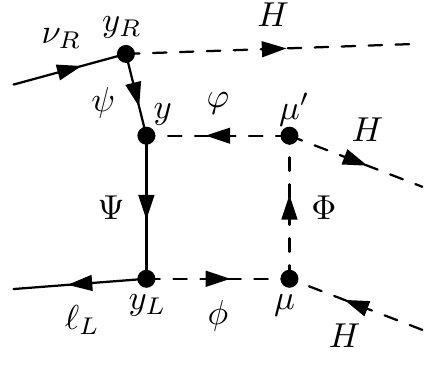}} & \multirow{2}{*}{} & \multirow{2}{*}{$\psi$} & \multirow{2}{*}{$\Psi$} & \multirow{2}{*}{$\phi$} & \multirow{2}{*}{$\varphi$} & \multirow{2}{*}{$\Phi$} & \multicolumn{2}{c|}{\text{Excluded} $\alpha$} & \multirow{2}{*}{\text{Dark Matter}}\\ \cline{8-9}
 &  &  &  &  &  &  & \text{Tree} & \text{Ma} & \\ \cline{2-10}
 & A & $\mathbf{2}_{-1}^F$ & $\mathbf{1}_{\alpha -1}^F$ & $\mathbf{2}_{\alpha }^S$ & $\mathbf{2}_{\alpha }^S$ & $\mathbf{1}_{\alpha +1}^S$ & \{\hl{$\pm 1$}\} & \{\hl{$\pm 1$}, $0$\} & $[\phi ,\varphi ,\Phi ]_{-1}^{\diamond}$, $[\Psi ,\phi ,\varphi ]_{1}^{\diamond}$\\ \cline{2-10}
 & B & $\mathbf{2}_{-1}^F$ & $\mathbf{1}_{\alpha -1}^F$ & $\mathbf{2}_{\alpha }^S$ & $\mathbf{2}_{\alpha }^S$ & $\mathbf{3}_{\alpha +1}^S$ & \{\hl{$-3$}, \hl{$\pm 1$}\} & \{\hl{$\pm 1$}\} & $[\phi ,\varphi ,\Phi ]_{-1}^{\diamond}$, $[\Psi ,\phi ,\varphi ,\Phi ]_{1}^{\diamond}$\\ \cline{2-10}
 & C & $\mathbf{2}_{-1}^F$ & $\mathbf{2}_{\alpha -1}^F$ & $\mathbf{1}_{\alpha }^S$ & $\mathbf{1}_{\alpha }^S$ & $\mathbf{2}_{\alpha +1}^S$ & \{\hl{$-2$}, \hl{$0$}\} & \{\hl{$-2$}, $0$\} & $[\Phi ]_{-2}^{\diamond}$, $[\phi ,\varphi ,\Phi ]_{0}^{\diamond}$\\ \cline{2-10}
 & D & $\mathbf{2}_{-1}^F$ & $\mathbf{2}_{\alpha -1}^F$ & $\mathbf{1}_{\alpha }^S$ & $\mathbf{3}_{\alpha }^S$ & $\mathbf{2}_{\alpha +1}^S$ & \{\hl{$\pm 2$}, \hl{$0$}\} & \{\hl{$-2$}, $0$\} & $[\varphi ,\Phi ]_{-2}$, $[\phi ,\varphi ,\Phi ]_{0}$\\ \cline{2-10}
 & E & $\mathbf{2}_{-1}^F$ & $\mathbf{2}_{\alpha -1}^F$ & $\mathbf{3}_{\alpha }^S$ & $\mathbf{1}_{\alpha }^S$ & $\mathbf{2}_{\alpha +1}^S$ & \{\hl{$\pm 2$}, \hl{$0$}\} & \{\hl{$-2$}, $0$\} & $[\phi ,\Phi ]_{-2}$, $[\phi ,\varphi ,\Phi ]_{0}$\\ \cline{2-10}
 & F & $\mathbf{2}_{-1}^F$ & $\mathbf{2}_{\alpha -1}^F$ & $\mathbf{3}_{\alpha }^S$ & $\mathbf{3}_{\alpha }^S$ & $\mathbf{2}_{\alpha +1}^S$ & \{\hl{$\pm 2$}, \hl{$0$}\} & \{\hl{$-2$}, $0$\} & $[\phi ,\varphi ,\Phi ]_{-2}^{\diamond}$, $[\phi ,\varphi ,\Phi ]_{0}^{\diamond}$\\ \cline{2-10}
 T8-5-3 & G & $\mathbf{2}_{-1}^F$ & $\mathbf{3}_{\alpha -1}^F$ & $\mathbf{2}_{\alpha }^S$ & $\mathbf{2}_{\alpha }^S$ & $\mathbf{1}_{\alpha +1}^S$ & \{\hl{$\pm 1$}, \hl{$3$}\} & \{\hl{$\pm 1$}\} & $[\phi ,\varphi ,\Phi ]_{-1}^{\diamond}$, $[\Psi ,\phi ,\varphi ]_{1}^{\diamond}$\\ \cline{2-10}
 & H & $\mathbf{2}_{-1}^F$ & $\mathbf{3}_{\alpha -1}^F$ & $\mathbf{2}_{\alpha }^S$ & $\mathbf{2}_{\alpha }^S$ & $\mathbf{3}_{\alpha +1}^S$ & \{\hl{$\pm 3$}, \hl{$\pm 1$}\} & \{\hl{$\pm 1$}, $0$\} & $[\phi ,\varphi ,\Phi ]_{-1}^{\diamond}$, $[\Psi ,\phi ,\varphi ,\Phi ]_{1}^{\diamond}$\\ \hline
\multicolumn{10}{|c|}{$(m_{\nu})_{\alpha\beta}/\langle H\rangle^3=-\frac{M_{\Psi _i}}{M_{\psi _j}}\mu  \mu ' {y_L}_{\alpha  i} y_{i j} {y_R}_{j \beta }I_{4}\left(M_{\phi },M_{\varphi },M_{\Phi },M_{\Psi _i}\right)$}\\ \hline\hline
\caption{\label{tab:figT8}
The finite diagrams generated from the topology T8. We list the possible quantum number assignments for the mediators, the expressions for the neutrino mass matrix, and the dark matter candidates. We also display the values of $\alpha$ excluded from the appearance of tree level and Ma diagrams, where $\varnothing$ and $\mathbb{U}$ denote empty set and universal set respectively.}
\end{longtable}

\clearpage

\section{\label{sec:app_mass}Functions involved in neutrino masses}

The integrals $I_n$, $J_n$ and $K_n$ appearing in the one-loop diagrams for Dirac neutrino masses are defined as follows :
\begin{equation}
\begin{aligned}
I_{2} (M_{A}, M_{B})\equiv & \int \frac{{\rm d}^{d} k}{(2\pi)^{d}i}\frac{1}{(k^{2} - M_{A}^{2})(k^{2} - M_{B}^{2})}\\
=&\frac{1}{(4\pi)^{2}} \left[\frac{2}{\epsilon} -\gamma_E +1+\ln(4\pi)-\ln
M_B^2+\frac{M_A^2\ln(\frac{M_B^2}{M_A^2})}{M_A^2-M_B^2}\right]\,,
\end{aligned}
\end{equation}

\begin{equation}
\label{eq:I3}
\begin{aligned}
I_{3} (M_{A}, M_{B}, M_{C})\equiv & \int \frac{{\rm d}^{d}
k}{(2\pi)^{d} i}\frac{1}{(k^{2} - M_{A}^{2})(k^{2} - M_{B}^{2})(k^{2}
- M_{C}^{2})} \\
=&\frac{1}{(4\pi)^{2}}\left[\frac{M_{A}^{2} \ln
\frac{M_{C}^{2}}{M_{A}^{2}}}{(M_{A}^{2} - M_{B}^{2})(M_{A}^{2} -
M_{C}^{2})}+\frac{M_{B}^{2} \ln \frac{M_{C}^{2}}{M_{B}^{2}}}{(M_{B}^{2} -
M_{A}^{2})(M_{B}^{2} - M_{C}^{2})}\right]\,,
  \end{aligned}
\end{equation}

\begin{equation}
\begin{aligned}
I_{4} (M_{A}, M_{B}, M_{C}, M_{D})\equiv & \int \frac{{\rm
d}^{d} k}{(2\pi)^{d} i}\frac{1}{(k^{2} - M_{A}^{2})(k^{2} -
M_{B}^{2})(k^{2} - M_{C}^{2})(k^{2} - M_{D}^{2})} \\
= & \frac{1}{(4\pi)^{2}}\left[\frac{M_{A}^{2} \ln
\frac{M_{D}^{2}}{M_{A}^{2}}}{(M_{A}^{2} - M_{B}^{2})(M_{A}^{2} -
M_{C}^{2})(M_{A}^{2} - M_{D}^{2})}\right.\\
& \hspace{1.3cm}
+\frac{M_{B}^{2} \ln \frac{M_{D}^{2}}{M_{B}^{2}}}{(M_{B}^{2} -
M_{A}^{2})(M_{B}^{2} - M_{C}^{2})(M_{B}^{2} - M_{D}^{2})}\\
& \hspace{1.3cm}
+\left.\frac{M_{C}^{2} \ln \frac{M_{D}^{2}}{M_{C}^{2}}}{(M_{C}^{2} -
M_{A}^{2})(M_{C}^{2} - M_{B}^{2})(M_{C}^{2} - M_{D}^{2})}\right]\,,
\end{aligned}
\end{equation}

\begin{equation}
\begin{aligned}
\hskip-0.2cm I_{5} (M_{A}, M_{B}, M_{C}, M_{D},
M_{E})\equiv & \int \frac{{\rm d}^{d} k}{(2\pi)^{d}i}\frac{1}{(k^{2} - M_{A}^{2})(k^{2} - M_{B}^{2})(k^{2} - M_{C}^{2})(k^{2}
- M_{D}^{2})(k^{2} - M_{E}^{2})} \\
= & \frac{1}{(4\pi)^{2}}\left[\frac{M_{A}^{2} \ln
\frac{M_{E}^{2}}{M_{A}^{2}}}{(M_{A}^{2} - M_{B}^{2})(M_{A}^{2} -
M_{C}^{2})(M_{A}^{2} - M_{D}^{2})(M_{A}^{2} - M_{E}^{2})}\right.\\
& \hspace{0.8cm}
+\frac{M_{B}^{2} \ln \frac{M_{E}^{2}}{M_{B}^{2}}}{(M_{B}^{2} -
M_{A}^{2})(M_{B}^{2} - M_{C}^{2})(M_{B}^{2} - M_{D}^{2})(M_{B}^{2} -
M_{E}^{2})}\\
& \hspace{0.8cm}
+\frac{M_{C}^{2} \ln \frac{M_{E}^{2}}{M_{C}^{2}}}{(M_{C}^{2} -
M_{A}^{2})(M_{C}^{2} - M_{B}^{2})(M_{C}^{2} - M_{D}^{2})(M_{C}^{2} -
M_{E}^{2})}\\
& \hspace{0.8cm}
+\left.\frac{M_{D}^{2} \ln \frac{M_{E}^{2}}{M_{D}^{2}}}{(M_{D}^{2} -
M_{A}^{2})(M_{D}^{2} - M_{B}^{2})(M_{D}^{2} - M_{C}^{2})(M_{D}^{2} -
M_{E}^{2})}\right]\,,
\end{aligned}
\end{equation}

\begin{equation}
\begin{aligned}
J_{3} (M_{A}, M_{B}, M_{C})\equiv&\int \frac{{\rm d}^{d}
k}{(2\pi)^{d}i}\frac{k^{2}}{(k^{2} - M_{A}^{2})(k^{2} -
M_{B}^{2})(k^{2} - M_{C}^{2})}\\
=&\frac{1}{(4\pi)^{2}}\left[ \frac{2}{\epsilon}-\gamma_{E}+1+\ln(4\pi)-\ln
M_{C}^{2}\right. \\
&\hspace{1cm}\left.
+\frac{M_A^4 \ln
\frac{M_C^2}{M_{A}^{2}}}{(M_A^2-M_B^2)(M_A^2-M_C^2)}+\frac{M_B^4 \ln
\frac{M_{C}^{2}}{M_B^2}}{(M_B^2-M_A^2)(M_B^2-M_C^2)}\right]\,,
\end{aligned}
\end{equation}

\begin{equation}
\begin{aligned}
J_{4} (M_{A}, M_{B}, M_{C}, M_{D})\equiv&\int \frac{{\rm
d}^{d} k}{(2\pi)^{d} i}\frac{k^{2}}{(k^{2} - M_{A}^{2})(k^{2} -
M_{B}^{2})(k^{2} - M_{C}^{2})(k^{2} - M_{D}^{2})} \\
=&\frac{1}{(4\pi)^{2}}\left[\frac{M_{A}^{4} \ln
\frac{M_{D}^{2}}{M_{A}^{2}}}{(M_{A}^{2} - M_{B}^{2})(M_{A}^{2} -
M_{C}^{2})(M_{A}^{2} - M_{D}^{2})}\right.\\
&\hspace{1.5cm}
+\frac{M_{B}^{4} \ln \frac{M_{D}^{2}}{M_{B}^{2}}}{(M_{B}^{2} -
M_{A}^{2})(M_{B}^{2} - M_{C}^{2})(M_{B}^{2} - M_{D}^{2})}\\
&\hspace{1.5cm}
\left.+\frac{M_{C}^{4} \ln \frac{M_{D}^{2}}{M_{C}^{2}}}{(M_{C}^{2} -
M_{A}^{2})(M_{C}^{2} - M_{B}^{2})(M_{C}^{2} - M_{D}^{2})}\right]\,,
\end{aligned}
\end{equation}

\begin{equation}
\begin{aligned}
\hskip-0.2cm J_{5} (M_{A}, M_{B}, M_{C}, M_{D},
M_{E})\equiv & \int \frac{{\rm d}^{d} k}{(2\pi)^{d}i}\frac{k^2}{(k^{2} - M_{A}^{2})(k^{2} - M_{B}^{2})(k^{2} -
M_{C}^{2})(k^{2} - M_{D}^{2})(k^{2} - M_{E}^{2})} \\
= & \frac{1}{(4\pi)^{2}}\left[\frac{M_{A}^{4} \ln
\frac{M_{E}^{2}}{M_{A}^{2}}}{(M_{A}^{2} - M_{B}^{2})(M_{A}^{2} -
M_{C}^{2})(M_{A}^{2} - M_{D}^{2})(M_{A}^{2} - M_{E}^{2})}\right. \\
& \hspace{0.8cm}
+\frac{M_{B}^{4} \ln \frac{M_{E}^{2}}{M_{B}^{2}}}{(M_{B}^{2} -
M_{A}^{2})(M_{B}^{2} - M_{C}^{2})(M_{B}^{2} - M_{D}^{2})(M_{B}^{2} -
M_{E}^{2})}\\
& \hspace{0.8cm}
+\frac{M_{C}^{4} \ln \frac{M_{E}^{2}}{M_{C}^{2}}}{(M_{C}^{2} -
M_{A}^{2})(M_{C}^{2} - M_{B}^{2})(M_{C}^{2} - M_{D}^{2})(M_{C}^{2} -
M_{E}^{2})}\\
& \hspace{0.8cm}
+\left.\frac{M_{D}^{4} \ln \frac{M_{E}^{2}}{M_{D}^{2}}}{(M_{D}^{2} -
M_{A}^{2})(M_{D}^{2} - M_{B}^{2})(M_{D}^{2} - M_{C}^{2})(M_{D}^{2} -
M_{E}^{2})}\right]\,,
\end{aligned}
\end{equation}

\begin{equation}
\begin{aligned}
K_{4} (M_{A}, M_{B}, M_{C}, M_{D})\equiv&\int \frac{{\rm
d}^{d} k}{(2\pi)^{d} i}\frac{k^{4}}{(k^{2} - M_{A}^{2})(k^{2} -
M_{B}^{2})(k^{2} - M_{C}^{2})(k^{2} - M_{D}^{2})} \\
=&\frac{1}{(4\pi)^{2}}\left[\frac{2}{\epsilon}-\gamma_{E}+1+\ln(4\pi)-\ln
M_{D}^{2}\right.\\
&\hspace{1.5cm}
+\frac{M_{A}^{6} \ln \frac{M_{D}^{2}}{M_{A}^{2}}}{(M_{A}^{2} -
M_{B}^{2})(M_{A}^{2} - M_{C}^{2})(M_{A}^{2} - M_{D}^{2})} \\
&\hspace{1.5cm}
+\frac{M_{B}^{6} \ln \frac{M_{D}^{2}}{M_{B}^{2}}}{(M_{B}^{2} -
M_{A}^{2})(M_{B}^{2} - M_{C}^{2})(M_{B}^{2} - M_{D}^{2})}\\
&\hspace{1.5cm}
\left.+\frac{M_{C}^{6} \ln \frac{M_{D}^{2}}{M_{C}^{2}}}{(M_{C}^{2} -
M_{A}^{2})(M_{C}^{2} - M_{B}^{2})(M_{C}^{2} - M_{D}^{2})}\right]\,,
\end{aligned}
\end{equation}

\begin{equation}
\begin{aligned}
\hskip-0.3cm K_{5} (M_{A}, M_{B}, M_{C}, M_{D},
M_{E})\equiv & \int \frac{{\rm d}^{d} k}{(2\pi)^{d}i}\frac{k^4}{(k^{2} - M_{A}^{2})(k^{2} - M_{B}^{2})(k^{2} -
M_{C}^{2})(k^{2} - M_{D}^{2})(k^{2} - M_{E}^{2})} \\
= & \frac{1}{(4\pi)^{2}}\left[\frac{M_{A}^{6} \ln
\frac{M_{E}^{2}}{M_{A}^{2}}}{(M_{A}^{2} - M_{B}^{2})(M_{A}^{2} -
M_{C}^{2})(M_{A}^{2} - M_{D}^{2})(M_{A}^{2} - M_{E}^{2})}\right. \\
& \hspace{0.8cm}
+\frac{M_{B}^{6} \ln \frac{M_{E}^{2}}{M_{B}^{2}}}{(M_{B}^{2} -
M_{A}^{2})(M_{B}^{2} - M_{C}^{2})(M_{B}^{2} - M_{D}^{2})(M_{B}^{2} -
M_{E}^{2})}\\
& \hspace{0.8cm}
+\frac{M_{C}^{6} \ln \frac{M_{E}^{2}}{M_{C}^{2}}}{(M_{C}^{2} -
M_{A}^{2})(M_{C}^{2} - M_{B}^{2})(M_{C}^{2} - M_{D}^{2})(M_{C}^{2} -
M_{E}^{2})}\\
& \hspace{0.8cm}
+\left.\frac{M_{D}^{6} \ln \frac{M_{E}^{2}}{M_{D}^{2}}}{(M_{D}^{2} -
M_{A}^{2})(M_{D}^{2} - M_{B}^{2})(M_{D}^{2} - M_{C}^{2})(M_{D}^{2} -
M_{E}^{2})}\right]\,,
\end{aligned}
\end{equation}
where $\epsilon=4-d$ is an infinitely small quantity, and $\gamma_E$ is the
Euler-Mascheroni constant, we can see that the functions $I_2, J_3$ and
$K_4$ are divergent, and the other functions are finite.

\end{appendix}

\providecommand{\href}[2]{#2}\begingroup\raggedright\endgroup

\end{document}